\documentclass[onecolumn,authoryear]{els-mrw} 

\usepackage{amsmath,amssymb,amsfonts,amsthm,makeidx,graphicx}
\usepackage{txfonts}
\usepackage{helvet}

\usepackage{times,graphicx}
\usepackage{authblk}

\usepackage{amsmath,amssymb,bm,mathtools}
\usepackage{physics}

\newcommand{\e}{\mathrm{e}}
\newcommand{\bk}{\mathbf{k}}
\newcommand{\bK}{\mathbf{K}}
\newcommand{\br}{\mathbf{r}}

\newcommand{\bq}{\mathbf{q}}

\newcommand{\bb}{\mathbf{b}}

\newcommand{\bx}{\mathbf{x}}
\newcommand{\by}{\mathbf{y}}

\newcommand{\bp}{\mathbf{p}}

\newcommand{\Tmat}{\mathcal{T}}

\begin{document}

\noindent
{\large\em Oxford Research Encyclopedia of Physics}$\,$ {\large (2026)}

\noindent
\hrulefill

\noindent
{\large\bf Nucleon Correlations in Unstable Nuclei via Knockout Reactions}\\

\noindent
Carlos A. Bertulani

\noindent
{\small
Department of Physics and Astronomy, East Texas A\&M University, Commerce, TX 75428, USA\\
Technische Universit\"at Darmstadt, Institut f\"ur Kernphysik, 64289 Darmstadt, Germany\\
ExtreMe Matter Institute EMMI, GSI Helmholtzzentrum f\"ur Schwerionenforschung GmbH, D-64291 Darmstadt, Germany\\
}

\noindent
{\small
Subject: Nuclear Physics

}

\vspace*{-0.06in}
\noindent
\hrulefill

\vspace*{0.10in}
\noindent
{\large\bf Summary and Keywords}

We review the physical ideas and reaction theory underlying one- and two-nucleon removal from fast projectiles, quasi-free proton-induced reactions such as $(p,2p)$ and $(p,pn)$, and the use of exclusive momentum and coincidence observables to diagnose correlations. Particular emphasis is placed on the distinction between independent-particle occupancy, long-range collective and pairing correlations, tensor-driven neutron-proton correlations, and short-range correlations generated by the repulsive core and noncentral components of the nuclear interaction. The discussion develops the overlap-function language, spectroscopic factors, eikonal stripping and diffraction, distorted-wave impulse approximation, factorization and spectral functions, center-of-mass and recoil effects, two-nucleon amplitudes, pair densities, and the relation between measured cross sections and ab initio or shell-model structure. The well-known reduction of experimental single-particle strength relative to simple shell-model expectations, its dependence on separation-energy asymmetry, and the continuing debate over reaction-model systematics are treated in detail. \\


\vspace*{-0.06in}
\noindent
\hrulefill

\vspace*{0.10in}
\noindent
{\large\bf Introduction}

The atomic nucleus is simultaneously a finite quantum liquid, a shell-structured fermionic system, and a strongly interacting many-body object.  The success of the independent-particle shell model might suggest that nucleons move almost independently in an average potential, yet essentially every quantitative observable reveals departures from that idealization.  Occupation numbers are depleted below the Fermi surface; nominally empty orbitals acquire finite occupation; pairing mixes configurations differing by two particles; collective quadrupole and octupole modes fragment single-particle strength; tensor forces preferentially correlate neutron-proton pairs; and the short-distance repulsion of the nucleon-nucleon interaction produces high-momentum components that cannot be represented by a single Slater determinant.  These effects are collectively known as nucleon correlations, but they occur on different length, momentum, and energy scales and therefore require different experimental filters (\cite{BohrMottelson1969,BohrMottelson1975,Mougey1976,RingSchuck1980,Kramer2001,Lapikas1993,Kelly1996,Bertulani2007,ObertelliSagawa2021}).

Unstable nuclei are particularly valuable laboratories because the balance among these mechanisms changes when the neutron-to-proton ratio, binding energy, continuum coupling, and shell gaps are varied.  Close to a drip line, a weakly bound nucleon can develop an extended wave function, the Fermi energy approaches the particle-emission threshold, and coupling to scattering states becomes inseparable from the internal structure.  At the same time, neutron and proton Fermi surfaces may be widely separated.  The resulting isospin asymmetry changes radial overlaps, pairing phase space, tensor-force effects, and the distribution of correlations between majority and minority species (\cite{Rios2009}).  Rare-isotope beams thus permit a controlled exploration of many-body physics under conditions unavailable near the valley of stability (\cite{BertulaniCantoHussein1993,HansenJensenJonson1995,AlKhaliliTostevin2002,HansenTostevin2003,AlKhaliliNunes2003,JensenRiisagerFedorovGarrido2004,BaranColonnaGrecoDiToro2005,BertulaniGade2010,Hagino2013,Moro2019,PASCHALIS2020135110,PhysRevC.100.064604,Alamanos2022,BertulaniBonaccorso2022,10.1093/ptep/ptaf119,HEBBORN2024138413}).

Knockout reactions are especially powerful in this setting.  In a fast collision, one or more nucleons can be removed on a time scale short compared with intrinsic nuclear motion.  If the reaction is sufficiently direct, the residue remembers the orbital angular momentum, separation energy, spatial distribution, and correlations of the removed nucleon or pair.  The method was developed into a major spectroscopic tool through high-energy one-nucleon removal from light targets and has subsequently expanded to two-nucleon removal, proton-induced quasi-free scattering, and increasingly exclusive measurements with tracking, calorimetry, neutron detection, and $\gamma$-ray arrays (\cite{HansenJonson1987,HansenTostevin2003,Tostevin2001,GadeTostevin2008,Aumann2013,JacobGade2014}).

The nuclear wavefunction is given by $
  \ket{\Psi_A}=\sum_\alpha C_\alpha\ket{\Phi_\alpha},
$
where $\ket{\Phi_\alpha}$ are wavefunctions of many-body configurations.  In an independent-particle picture one coefficient dominates.  Correlations are encoded in the coherent admixture of other configurations.  
In second quantization, let
$\psi^\dagger(\mathbf r\sigma\tau)$ and
$\psi(\mathbf r\sigma\tau)$ denote, respectively, the fermionic
field operators that create and annihilate a nucleon at position
$\mathbf r$, with spin projection $\sigma$ and isospin projection
$\tau$. They obey the canonical anticommutation relation
\begin{equation}
 \left\{
 \psi(\mathbf r\sigma\tau),
 \psi^\dagger(\mathbf r'\sigma'\tau')
 \right\}
 =
 \delta(\mathbf r-\mathbf r')
 \delta_{\sigma\sigma'}
 \delta_{\tau\tau'} .
\end{equation}
Introducing the compact notation
$x=(\mathbf r,\sigma,\tau)$, the one-body density matrix of an
$A$-nucleon state $|\Psi_A\rangle$ is
\begin{equation}
 \rho(x,x')
 =
 \langle\Psi_A|
 \psi^\dagger(x')\psi(x)
 |\Psi_A\rangle .
\end{equation}
The natural orbitals $\phi_\nu(x)$ are the eigenfunctions of this
density matrix,
\begin{equation}
 \int dx'\,\rho(x,x')\phi_\nu(x')
 =
 n_\nu\phi_\nu(x),
\end{equation}
where the eigenvalues $n_\nu$ are the corresponding occupation
numbers. For fermions, $0\le n_\nu\le1$ and
$\sum_\nu n_\nu=A$. In an independent-particle Slater determinant
the occupation numbers are exactly zero or one, whereas nucleon-nucleon correlations generate fractional occupations and redistribute
single-particle strength across the Fermi surface.
Likewise, the two-body density
\begin{equation}
 \rho^{(2)}(1,2;1',2')=\bra{\Psi_A}\psi^\dagger(1')\psi^\dagger(2')\psi(2)\psi(1)\ket{\Psi_A}
\end{equation}
contains information that cannot be reconstructed from $\rho$ alone.  Two-nucleon knockout is potentially sensitive to precisely this irreducible pair information.

The challenge of a theoretical interpretation of an experiment is that a measured cross section is never directly proportional to a bare occupation number.  It is a convolution of nuclear structure with a reaction mechanism.  A schematic distorted-wave Born approximation (DWBA) expression for an exclusive cross section may be written as
\begin{equation}
  \dd\sigma \sim \left|\bra{\chi_f^{(-)}\Psi_{A-1}^{f}}\hat{V}_{\rm tr}\ket{\chi_i^{(+)}\Psi_A}\right|^2 \dd P,
\end{equation}
where $\chi_i$ and $\chi_f$ describe relative motion, $\hat V_{\rm tr}$ is the transition operator, and $\dd P$ is final-state phase space.  Different reaction theories make different controlled approximations to this matrix element.  The scientific value of knockout reactions depends on identifying observables for which the reaction mechanism (via descriptions of $\chi_i$, $\chi_f$, $\hat V_{\rm tr}$, and $\dd P$) is well understood and the desired structural information is not washed out by unobserved degrees of freedom. Further complications arise if higher-order processes (e.g., multiple nucleon-nucleon collisions) play an important role (\cite{BERTULANI2023138250}).

\vspace*{0.06in}
\noindent
{\large\bf Correlated overlap functions}

The quantity most directly connected with one-nucleon removal is the overlap between the projectile and a specific state of the $A-1$ residue,
\begin{equation}
 I_{\alpha}(\br,\sigma,\tau)
 =\sqrt{A}\,\left\langle{\Psi_{A-1}^{\alpha}}\Big|\Psi_A(\br,\sigma,\tau;2,\ldots,A)\right\rangle. \label{overlap}
\end{equation}
Its norm defines a spectroscopic factor,
\begin{equation}
 S_\alpha=\int \dd^3r\sum_{\sigma\tau}|I_\alpha(\br,\sigma,\tau)|^2.
\end{equation}
In a simple shell-model limit $I_\alpha$ is proportional to a normalized single-particle wave function and $S_\alpha$ reduces to the corresponding independent-particle occupancy or shell-model spectroscopic strength.  In the exact many-body problem, however, the overlap is neither generally normalized to unity nor described by a Woods-Saxon orbital or a similar single-particle wavefunction.  Correlations distribute the strength associated with a nominal orbital over many final states and over a broad continuum.

For weakly bound unstable nuclei the asymptotic form of the overlap is itself essential.  For a neutron with separation energy $S_n$, the overlap function, Eq. \eqref{overlap}, becomes
\begin{equation}
 I_{lj}(r)\xrightarrow[r\to\infty]{} C_{lj}\frac{\e^{-\kappa r}}{r},\qquad
 \kappa=\frac{\sqrt{2\mu S_n}}{\hbar},
\end{equation}
where $C_{lj}$ is an asymptotic normalization coefficient.  For protons the asymptotic function is Coulomb modified.  Because fast removal on light targets is surface localized, the calculated cross section can depend strongly on this tail.  This is one reason why a separation-energy-matched single-particle potential is normally used in eikonal calculations.

A complementary description of nucleon correlations is provided by the one-body Green function, or single-particle propagator. Whereas the one-body density matrix characterizes the occupation of single-particle orbitals in the nuclear ground state, the Green function describes the propagation of a nucleon added to the nucleus, or of a hole produced by removing one. It therefore provides a natural connection between many-body correlations and nucleon-addition or nucleon-removal experiments (\cite{Aumann2000,Bazin2003,Rohe2004,Sauvan2004,Gade2004,Yoneda2006,Kobayashi2012,Holl2019,PhysRevC.93.044607,PaninAumannBertulani2021,Panin2016,Panin2019,Panin2026}).
Let $|\Psi_A^0\rangle$ denote the correlated ground state of an $A$-nucleon nucleus with energy $E_A^0$. Let $a_\alpha^\dagger$ and $a_\alpha$ create and annihilate, respectively, a nucleon in the single-particle state $\alpha$. The time-ordered one-body Green function is (\cite{DickhoffBarbieri2004,BarbieriCarbone2017})
\begin{equation}
G_{\alpha\beta}(t-t')
=
-i\langle\Psi_A^0|
T\left[a_\alpha(t)a_\beta^\dagger(t')\right]
|\Psi_A^0\rangle,
\label{eq:green_time}
\end{equation}
where $T$ is the time-ordering operator. Its Fourier transform is
\begin{equation}
G_{\alpha\beta}(E)
=
\int_{-\infty}^{\infty} d\tau\,
e^{iE\tau}G_{\alpha\beta}(\tau),
\qquad \tau=t-t'.
\label{eq:green_energy}
\end{equation}

The physical content of the propagator is most transparent in its Lehmann representation (\cite{Soma2020}). Inserting complete sets of eigenstates of the $(A+1)$- and $(A-1)$-body systems gives 
\begin{align}
G_{\alpha\beta}(E)
=
\sum_n
\frac{
\langle\Psi_A^0|a_\alpha|\Psi_{A+1}^n\rangle
\langle\Psi_{A+1}^n|a_\beta^\dagger|\Psi_A^0\rangle
}
{E-(E_{A+1}^n-E_A^0)+i\eta}
+
\sum_k
\frac{
\langle\Psi_A^0|a_\beta^\dagger|\Psi_{A-1}^k\rangle
\langle\Psi_{A-1}^k|a_\alpha|\Psi_A^0\rangle
}
{E+(E_{A-1}^k-E_A^0)-i\eta}.
\label{eq:lehmann}
\end{align}
The first term describes particle propagation: a nucleon is added to the $A$-body ground state and the system explores eigenstates $|\Psi_{A+1}^n\rangle$. The second term describes hole propagation: a nucleon is removed and the residual system is left in an eigenstate $|\Psi_{A-1}^k\rangle$.
The overlap amplitude for removing a nucleon from orbital $\alpha$ and leaving the residual nucleus in the state $|\Psi_{A-1}^k\rangle$ is
\begin{equation}
\mathcal{A}_{\alpha}^{(k)}
=
\langle\Psi_{A-1}^k|a_\alpha|\Psi_A^0\rangle,
\end{equation}
and the corresponding spectroscopic strength is
\begin{equation}
S_{\alpha}^{(k)}
=
\left|
\langle\Psi_{A-1}^k|a_\alpha|\Psi_A^0\rangle
\right|^2.
\label{eq:spectroscopic_strength}
\end{equation}

For an independent-particle system, suppose that the orbital $\alpha$ has the unperturbed single-particle energy $\varepsilon_\alpha^{(0)}$. The corresponding propagator is
\begin{equation}
G_\alpha^{(0)}(E)
=
\frac{1}{E-\varepsilon_\alpha^{(0)}+i\eta}.
\label{eq:free_green}
\end{equation}
It has a single pole at $E=\varepsilon_\alpha^{(0)}$, meaning that all of the single-particle strength associated with the orbital $\alpha$ is concentrated in one state.
In an interacting nucleus, the propagating nucleon couples to more complicated configurations, such as particle-hole excitations, collective surface vibrations, pairing modes, and higher-order many-particle--many-hole configurations. These effects are summarized by the irreducible self-energy $\Sigma(E)$. The full propagator satisfies the Dyson equation (\cite{Dickhoff2017})
\begin{equation}
G(E)=G^{(0)}(E)+G^{(0)}(E)\Sigma(E)G(E).
\label{eq:dyson}
\end{equation}
If the propagator and self-energy are approximately diagonal in the chosen single-particle basis, one may write
\begin{equation}
G_\alpha(E)
=
\frac{1}{E-\varepsilon_\alpha^{(0)}-\Sigma_\alpha(E)}.
\label{eq:green_selfenergy}
\end{equation}
In general, the self-energy is complex,
$
\Sigma_\alpha(E)
=
\operatorname{Re}\Sigma_\alpha(E)
+i\,\operatorname{Im}\Sigma_\alpha(E).
$
Its real part shifts the location of the single-particle level, whereas its imaginary part describes the coupling to configurations into which the single-particle excitation can spread. Above open-channel thresholds, the imaginary part is also associated with a finite lifetime or decay width.

A quasiparticle (quasihole) state is an eigenstate of the interacting
$A+1$ $(A-1)$ nucleon system that retains a substantial overlap
with the state obtained by adding (removing) a nucleon in a specified
single-particle orbital $\alpha$ to (from) the correlated $A$-body
ground state. Thus, a quasiparticle or quasihole is not a pure
single-particle excitation, but rather a single-particle or
single-hole configuration dressed by many-body correlations. The
remaining single-particle strength is fragmented among other
eigenstates of the neighboring nucleus. For an isolated bound quasiparticle or quasihole state, the pole energy $E_\alpha$ satisfies
$
E_\alpha-\varepsilon_\alpha^{(0)}
-\operatorname{Re}\Sigma_\alpha(E_\alpha)=0.
\label{eq:pole_condition}
$
Expanding the denominator of Eq.~(\ref{eq:green_selfenergy}) around $E=E_\alpha$ gives
\begin{align}
E-\varepsilon_\alpha^{(0)}-\Sigma_\alpha(E)
\simeq
(E-E_\alpha)
\left[
1-
\left.
\frac{\partial\Sigma_\alpha(E)}{\partial E}
\right|_{E=E_\alpha}
\right].
\end{align}
Hence, close to the pole,
\begin{equation}
G_\alpha(E)
\simeq
\frac{Z_\alpha}{E-E_\alpha+i\eta}
+G_\alpha^{\mathrm{background}}(E),
\label{eq:pole_green}
\end{equation}
where
\begin{equation}
Z_\alpha
=
\left[
1-
\left.
\frac{\partial\Sigma_\alpha(E)}{\partial E}
\right|_{E=E_\alpha}
\right]^{-1}
\label{eq:quasiparticle_residue}
\end{equation}
is the quasiparticle residue and $G_\alpha^{\mathrm{background}}(E)$ is all remaining strength.

The quantity $Z_\alpha$ measures how much single-particle character is carried by that pole. In the independent-particle limit, the self-energy has no relevant energy dependence and $Z_\alpha=1$. In a correlated system, the energy dependence of $\Sigma_\alpha(E)$ generally leads to $Z_\alpha<1$. The missing strength is not destroyed; it is redistributed over other eigenstates of the neighboring nucleus. This redistribution is referred to as the fragmentation of single-particle strength.

The same physics is expressed by the hole spectral function (\cite{Frick2004}),
\begin{equation}
S_h(\alpha,E)
=
\sum_k
\left|
\langle\Psi_{A-1}^k|a_\alpha|\Psi_A^0\rangle
\right|^2
\delta\!\left(E-[E_A^0-E_{A-1}^k]\right).
\label{eq:hole_spectral}
\end{equation}
Its integral up to the Fermi energy gives the occupation of orbital $\alpha$,
\begin{equation}
n_\alpha
=
\int_{-\infty}^{E_F} S_h(\alpha,E)\,dE.
\label{eq:occupation_spectral}
\end{equation}
Thus the occupation numbers obtained from the one-body density matrix and the fragmentation described by the Green function are two manifestations of the same many-body correlations.

This formalism provides a microscopic interpretation of nucleon-removal reactions. In a simple factorized description, the theoretical cross section for removing a nucleon from orbital $\alpha$ and populating a specific state $k$ of the residual nucleus is written schematically as
\begin{equation}
\sigma_{\mathrm{th}}^{(k)}
=
S_{\alpha}^{(k)}\,\sigma_{\mathrm{sp}}^{(k)},
\label{eq:knockout_sf}
\end{equation}
where $\sigma_{\mathrm{sp}}^{(k)}$ is the calculated single-particle removal cross section for unit spectroscopic strength. Correlations that fragment the single-particle strength therefore reduce the strength carried by any one final state and can lead to measured exclusive knockout cross sections smaller than independent-particle expectations.
The sequence may be summarized schematically as correlations $\rightarrow$ an energy-dependent self-energy $\Sigma(E) \rightarrow Z_\alpha<1$ and fragmentation of single-particle strength $\rightarrow$ reduced strength in individual exclusive knockout channels

One should nevertheless be careful when interpreting spectroscopic factors as direct observables. Their numerical values depend on the resolution scale, the nuclear Hamiltonian, the many-body truncation, and the operators used to define the single-particle degrees of freedom. In addition, their extraction from measured knockout cross sections depends on the reaction model and therefore on quantities such as optical potentials, nucleon--target interactions, bound-state wave functions, and approximations in the reaction dynamics (\cite{Hebborn2023,PhysRevC.105.024613,GOMEZRAMOS2023138284,BERTULANI2023138250,PhysRevC.98.034610,Capel2022}). Within a consistently defined structure and reaction framework, however, the energy dependence of the self-energy and the associated fragmentation of the spectral function provide a microscopic interpretation of the depletion and redistribution of single-particle strength observed in nucleon-removal reactions  (\cite{DickhoffBarbieri2004,BarbieriDickhoff2009,Duguet2015}).

It is useful to separate correlations by characteristic scales.  Long-range correlations (LRC) arise from low-energy collective vibrations, deformation, pairing, and configuration mixing across nearby shell gaps.  They strongly fragment strength near the Fermi surface.  Short-range correlations (SRC) arise from the hard central and tensor components of the interaction and transfer strength to large missing energy and momentum.  In uniform matter the momentum distribution may be written schematically
\begin{equation}
 n(k)=n_{\rm MF}(k)+\delta n_{\rm LRC}(k)+\delta n_{\rm SRC}(k),
\end{equation}
where $\delta n_{\rm SRC}$ generates a tail extending far above the Fermi momentum.  In finite nuclei the decomposition is not unique, but it is physically useful because different reactions sample different domains (\cite{DickhoffBarbieri2004,FrankfurtStrikman1988,Weinstein2011,Arrington2012,Sargsian2014,Hen2017,Duer2018,Ryckebusch2019}).

\vspace*{0.12in}
\noindent
{\large\bf Fast one-nucleon removal on light nuclear targets}

The classic rare-isotope knockout experiment accelerates a projectile $P=c+N$ to tens or hundreds of MeV per nucleon and directs it onto a light target, commonly $^9$Be or $^{12}$C.  The collision removes nucleon $N$ while the core $c$ continues near the beam velocity.  At sufficiently high energy the sudden and eikonal approximations become effective.  The projectile-target relative trajectory is approximately straight, with impact parameter $\bb$.  Each projectile constituent accumulates an eikonal phase and is represented by an $S$ matrix,
\begin{equation}
 S_i(b)=\exp\left[i\chi_i(b)\right],\qquad
 \chi_i(b)=-\frac{1}{\hbar v}\int_{-\infty}^{\infty}U_i\left(\sqrt{b^2+z^2}\right)\dd z. \label{eikphas}
\end{equation}
These $S$ matrices may be obtained from optical potentials or from Glauber folding of nucleon-nucleon profile functions with projectile and target densities (\cite{Glauber1959,Bertulani2003,HansenTostevin2003,BertulaniHansen2004,BertulaniDeConti2010,LiPeiPang2026}).

Two leading mechanisms contribute.  In stripping, or inelastic breakup, the removed nucleon interacts nonelastically with the target while the residue survives elastically.  For a normalized single-particle orbital $\phi_{jm}$, the stripping cross section has the form (\cite{Hencken1996,Bertulani2003})
\begin{equation}
 \sigma_{\rm str}=\frac{1}{2j+1}\sum_m\int\dd^2b\int\dd^3r\,
 |S_c(b_c)|^2\left[1-|S_N(b_N)|^2\right]|\phi_{jm}(\br)|^2.
\end{equation}
The factor $|S_c|^2$ imposes residue survival, while $1-|S_N|^2$ requires removal of the nucleon.  This product localizes the reaction to the projectile surface: deeply interior nucleons are geometrically shadowed because configurations in which the core strongly overlaps the target are absorbed.

In diffraction, or elastic breakup, both the core and nucleon survive elastically but dissociate.  A compact expression is (\cite{Hencken1996,Bertulani2003})
\begin{align}
 \sigma_{\rm diff}=\frac{1}{2j+1}\sum_m\int\dd^2b
 \left\langle |S_cS_N|^2\right\rangle_m
 -\frac{1}{2j+1}\sum_{mm'}\int\dd^2b
 \left|\left\langle\phi_{jm'}|S_cS_N|\phi_{jm}\right\rangle\right|^2,
\end{align}
where the second term removes events in which the projectile remains bound.  Coulomb breakup may also contribute, particularly for weakly bound projectiles and high-$Z$ targets, but is usually small for the light targets chosen for nuclear knockout spectroscopy.

For a transition to residue state $\alpha$, the theoretical partial cross section is conventionally factorized as
\begin{equation}
 \sigma_{\rm th}(\alpha)=\left(\frac{A}{A-1}\right)^N C^2S(\alpha)\,\sigma_{\rm sp}(nlj,S_N^{\rm eff}),
\end{equation}
where $C^2S$ is the shell-model spectroscopic factor, the prefactor is a center-of-mass correction often used with harmonic-oscillator shell-model wave functions, $N$ is the major oscillator quantum number, and
$
 S_N^{\rm eff}=S_N+E_x(\alpha)
$
is the effective separation energy.  The single-particle cross section
$
 \sigma_{\rm sp}=\sigma_{\rm str}+\sigma_{\rm diff}
$
contains the reaction dynamics and the geometry of the bound-state orbital.

The inclusive theoretical cross section sums all bound residue final states,
$
 \sigma_{\rm th}^{\rm inc}=\sum_{\alpha\in\mathrm{bound}}\sigma_{\rm th}(\alpha).
$
Comparison with experiment is summarized by the reduction factor
$
 R_s={\sigma_{\rm exp}}/{\sigma_{\rm th}}.
$
Values $R_s<1$ have often been interpreted as evidence that correlations deplete independent-particle strength.  Systematic surveys found a striking dependence on the neutron-proton separation-energy asymmetry, suggesting stronger reduction for removal of a deeply bound minority nucleon from an asymmetric projectile (\cite{Gade2008,TostevinGade2014}).  This observation stimulated extensive theoretical discussion because transfer and quasifree-scattering analyses have sometimes inferred a weaker asymmetry dependence (\cite{Kay2013,Atar2018,GomezRamosMoro2018,Aumann2021}).

The interpretation of $R_s$ requires care.  It contains both structure and reaction-model errors  (\cite{Aumann2021,RODRIGUEZSANCHEZ2024138559}).  If the single-particle radial wave function is too compact, if the optical $S$ matrices over-absorb, if residue feeding is incomplete, or if unbound strength is omitted, $R_s$ can change even without a change in intrinsic correlations (\cite{PhysRevC.73.044608,PhysRevLett.131.212503,XIE2023137800}).  Conversely, modern many-body calculations predict genuine quenching from configuration mixing and short-range physics.  The scientifically meaningful task is therefore not to identify $R_s$ with a universal spectroscopic factor, but to establish consistency among multiple reactions and structure models at a specified resolution.

An advantage of fast knockout is that the longitudinal momentum distribution of the residue retains information about the removed nucleon's orbital angular momentum.  In the sudden limit, the distribution is closely related to a Fourier transform of the projectile overlap, modified by the impact-parameter-dependent survival and removal probabilities (\cite{BertulaniMcVoy1992}).  Schematically,
\begin{equation}
 \frac{\dd\sigma}{\dd k_z}\propto\sum_m\int\dd^2b\left|\int\dd^3r\,
 \e^{-ik_z z}\,\mathcal{W}(\bb,\br)\phi_{lm}(\br)\right|^2,
\end{equation}
where $\mathcal{W}$ denotes the reaction weighting.  Low-$l$ weakly bound orbitals have extended coordinate-space tails and therefore narrow momentum distributions, while more localized or higher-$l$ configurations typically generate broader structures.
The connection is not a simple free Fourier transform (\cite{BertulaniHansen2004}).  Strong absorption samples the nuclear surface and breaks spherical symmetry.  The magnetic substates are weighted differently, and the transverse momentum distribution is broadened by deflection and recoil.  Nevertheless, calculated parallel-momentum shapes have proven robust enough to assign $l$ values in many rare isotopes.  When $\gamma$-rays identify individual residue states, the method becomes a state-resolved spectroscopic probe.

Correlations enter momentum distributions in several ways.  First, configuration mixing changes which $nlj$ orbitals contribute to a given final state.  Second, weak binding and continuum coupling modify radial tails.  Third, short-range correlations create high-momentum components, although standard heavy-ion one-nucleon removal is strongly surface dominated and therefore not optimized for observing the most compact SRC configurations.  Fourth, in two-nucleon removal the pair's total orbital angular momentum controls the residue momentum width and shape in a way that can expose the angular-momentum coupling of correlated configurations (\cite{Simpson2009,SimpsonTostevin2010}).

\vspace*{0.12in}
\noindent
{\large\bf Two-nucleon removal and spatial pair correlations}

For a pair of active orbitals $\beta_1=(n_1l_1j_1)$ and $\beta_2=(n_2l_2j_2)$ coupled to total angular momentum $LS$, the two-nucleon overlap can be expanded
\begin{equation}
 \Psi_i^{(F)}(1,2)=\sum_{\alpha}C_{\alpha}\left[\phi_{\beta_1}(1)\otimes\phi_{\beta_2}(2)\right]_{LS\alpha},
\end{equation}
where $C_{\alpha}$ are two-nucleon amplitudes.  The distribution over $LS$ is particularly important for the residue parallel momentum distribution.  Thus an exclusive two-nucleon knockout measurement can test more than the total pair-removal strength: it can constrain the internal angular-momentum composition of the overlap (\cite{SimpsonTostevin2010,Longfellow2020}).

Direct two-nucleon removal extends the one-body logic to the two-body density.  At high energy on a light target, two nucleons may be stripped, one may be stripped while the other diffracts, or both may undergo elastic breakup.  The double-stripping contribution for a projectile state $J_i$ and residue final state $J_f$ has the generic structure
\begin{equation}
 \sigma_{\rm str}^{(2)}=\frac{1}{2J_i+1}\sum_{M_iM_f}\int\dd^2b\,
 \bra{\Psi_{J_iM_i}^{(J_fM_f)}}
 |S_c|^2(1-|S_1|^2)(1-|S_2|^2)
 \ket{\Psi_{J_iM_i}^{(J_fM_f)}}.
\end{equation}
The operator demands that both nucleons encounter the target while the residue survives.  It therefore samples the joint probability that the two nucleons are close to the same region of the projectile surface.  This is the geometric origin of sensitivity to spatial pair correlations.

We define the pair density for a transition to residue state $f$ as
\begin{equation}
 \rho_f^{(2)}(\br_1,\br_2)=\sum_{M_iM_f}|\Psi_{J_iM_i}^{(fJ_fM_f)}(\br_1,\br_2)|^2.
\end{equation}
A useful visualization is the probability at fixed radii as a function of opening angle $\omega$, $
 \cos\omega=\hat{\br}_1\cdot\hat{\br}_2$.
Coherent two-nucleon amplitudes can strongly enhance configurations with small opening angle, thereby increasing the probability that both nucleons overlap the target in a peripheral collision.  The cross section is consequently sensitive to interference among shell-model components, not merely to the incoherent sum of their occupations (\cite{SimpsonTostevin2011}).

For like nucleons, antisymmetry constrains the allowed spin, isospin, and spatial combinations.  For neutron-proton removal both $T=0$ and $T=1$ pairs are possible, allowing access to isoscalar correlations that are absent for identical pairs.  Studies of $^{12}$C found that measured $np$ removal was enhanced relative to calculations based on standard shell-model overlaps, whereas like-pair channels were better reproduced, suggesting missing $T=0$ spatial correlations (\cite{SimpsonTostevin2011}).  Subsequent calculations with no-core shell-model overlaps generated by chiral two- plus three-nucleon interactions showed that $np$ removal to low-lying $T=0$ states can indeed be significantly sensitive to the microscopic interaction and multi-$\hbar\omega$ components (\cite{SimpsonNavratil2012}).

Pair removal is also a spectroscopy tool at the limits of stability.  Removing two like nucleons can reach nuclei inaccessible by one-nucleon removal while populating states according to the parent two-nucleon amplitudes.  Residue momentum distributions carry information about the pair's total orbital angular momentum and hence about final-state spin.  This was demonstrated in reactions leading to neutron-rich and proton-rich magnesium-region nuclei (\cite{Simpson2009PRL}).  More recent measurements have exploited exclusive momentum distributions to test the composition of states in proton-rich $sd$-shell systems (\cite{Longfellow2020}).

An important distinction must be maintained between ``correlated removal'' and ``removal of correlated nucleons.''  Two nucleons may be removed in a direct event even if the initial wave function is approximated by a shell-model pair.  Conversely, an initial short-range-correlated pair may not dominate an inclusive two-nucleon removal cross section because the reaction is surface weighted and because sequential or indirect mechanisms may contribute.  A credible claim about correlations therefore requires an observable demonstrated theoretically to vary with the correlation component under study.

\vspace*{0.06in}
\noindent
{\large\bf Quasi-free scattering in inverse kinematics: $(p,2p)$ and $(p,pn)$}

Hydrogen-induced quasi-free scattering has become a central tool for rare-isotope spectroscopy.  In inverse kinematics a fast projectile $A$ collides with a proton target and a bound nucleon is knocked out.  Proton removal is observed as
$
 A(p,2p)B,
$
while neutron removal proceeds through
$
 A(p,pn)B.
$
At sufficiently high energy and momentum transfer, the elementary $pN$ collision is localized on a scale smaller than the nucleus, motivating an impulse approximation in which the projectile nucleon is struck quasi-freely (\cite{Aumann2013,Atar2018,WakasaOgataNoro2017,10.1093/ptep/ptaf119}).

In plane-wave impulse approximation, the transition amplitude can be factorized schematically as
\begin{equation}
 \Tmat_{fi}\approx t_{pN}(\bq)\,\widetilde I_\alpha(\bk_m),
\end{equation}
where $t_{pN}$ is the free or effective proton-nucleon scattering amplitude and
$
 \bk_m=\bk_1+\bk_2-\bk_0
$
is a missing momentum defined by the measured initial and final nucleon momenta in an appropriate frame.  The corresponding exclusive cross section has the form
\begin{equation}
 \frac{\dd^3\sigma}{\dd E_1\dd\Omega_1\dd\Omega_2}
 =K\,\frac{\dd\sigma_{pN}}{\dd\Omega}\,S(E_m,\bk_m),
\end{equation}
where $K$ is a kinematic factor and $S(E_m,\bk_m)$ is a spectral function defined in Eq. \eqref{eq:hole_spectral}.  The missing energy $E_m$ identifies the energy required to remove the nucleon and leave the residue in a particular configuration.
A sufficiently exclusive knockout experiment is, in principle, a direct probe of the correlated distribution of removal strength in momentum and energy.  In practice, distortion of incoming and outgoing waves, recoil, finite-range effects, off-shell ambiguities, and final-state interactions must be included (\cite{Aumann2021}).

Distorted-wave impulse approximation replaces plane waves by optical-model scattering waves,
\begin{equation}
 \Tmat_{fi}^{\rm DWIA}=\left\langle
 \chi_1^{(-)}\chi_2^{(-)}\Phi_B
 \left|t_{pN}\right|
 \chi_0^{(+)}\Phi_A
 \right\rangle.
\end{equation}
After suitable approximations, one often writes
\begin{equation}
 \sigma_{\rm th}=C^2S\,\sigma_{\rm sp}^{\rm DWIA}.
\end{equation}
Because the proton target is elementary and the outgoing nucleons can be detected, $(p,2p)$ and $(p,pn)$ offer complementary systematics to heavy-ion removal.  They can be less strongly surface dominated, particularly at several hundred MeV per nucleon, and complete kinematics can suppress ambiguities from indirect feeding.

The reaction is not perfectly transparent.  Initial- and final-state interactions attenuate trajectories through the nuclear interior.  In semiclassical language the transparency may be represented by
\begin{equation}
 \mathcal{P}(\br)=P_{\rm in}(\br)P_1^{\rm out}(\br)P_2^{\rm out}(\br),
\end{equation}
with, for example,
\begin{equation}
 P_{\rm in}(\br)\simeq\exp\left[-\int_{-\infty}^{z}\sigma_{pN}^{\rm tot}\rho(\bb,z')\dd z'\right].
\end{equation}
Thus even quasifree scattering has a radial filter.  Comparing its radial sensitivity with that of carbon- or beryllium-induced removal is an important way to diagnose whether apparent quenching originates in structure or reaction dynamics.

Modern experiments use thick liquid-hydrogen targets, vertex tracking, proton detectors, neutron arrays, $\gamma$-ray spectroscopy, and large-acceptance residue spectrometers.  These systems can reconstruct missing momentum, opening angles, excitation energies, and correlations among outgoing particles.  The development transforms knockout from an inclusive counting method into multidimensional spectroscopy (\cite{Aumann2021,10.1093/ptep/ptaf119,BertulaniDoornenbalObertelliUesaka2026}).

\vspace*{0.06in}
\noindent
{\large\bf Spectroscopic-factor quenching and separation-energy asymmetry}

One of the most debated results from intermediate-energy one-nucleon removal is the systematic reduction of measured cross sections relative to shell-model plus eikonal predictions.  To organize neutron and proton removal on a common axis, an asymmetry variable is often defined.  For neutron removal one may take
$
 \Delta S=S_n-S_p,
$
and for proton removal
$
 \Delta S=S_p-S_n,
$
with modifications for transitions to excited residue states.  Positive $\Delta S$ then corresponds roughly to removal of a more deeply bound nucleon species.  Surveys found $R_s$ approaching unity for weakly bound nucleons but falling substantially for deeply bound nucleons in strongly asymmetric systems (\cite{Gade2008,TostevinGade2014}).

A many-body explanation is plausible.  In neutron-rich matter the minority protons can couple strongly through tensor-dominated $np$ correlations to the abundant neutrons, producing enhanced high-momentum proton components.  The low-energy quasihole strength for deeply bound protons could therefore be more depleted.  Self-consistent Green-function calculations and other correlated methods predict nontrivial isospin dependence of spectroscopic strength, though generally the predicted trends depend on Hamiltonian, model space, and definition of the quasiparticle strength.

However, the magnitude of the experimental trend has been questioned.  Transfer reactions analyzed with consistent methods have not always shown comparable asymmetry dependence (\cite{Kay2013}).  Quasifree $(p,2p)$ measurements of oxygen isotopes have provided important cross-checks and stimulated renewed scrutiny of reaction theory (\cite{Atar2018}).  Calculations beyond the simplest eikonal approximations, including dynamical effects and improved optical interactions, can modify extracted reduction factors (\cite{GomezRamosMoro2018}).

There is also a formal issue: spectroscopic factors are not observables in the strict field-theoretic sense.  A unitary transformation of the Hamiltonian can redistribute strength between wave functions and operators without changing exact cross sections.  Therefore a quoted $C^2S$ is meaningful only together with a specified Hamiltonian, resolution scale, basis, and reaction operator (\cite{FurnstahlHammer2001,FurnstahlSchwenk2010,Duguet2015}).  This does not make spectroscopic factors useless.  It means that comparisons must be internally consistent, much as parton distributions in hadronic physics depend on factorization scheme and scale while remaining extraordinarily useful.

A productive strategy is to compare several reactions on the same nuclei using structure inputs generated consistently.  If heavy-ion removal, proton-induced quasifree scattering, transfer, and electron scattering all require similar depletion of a given overlap, the structural interpretation becomes much stronger.  If they disagree, the pattern can expose missing reaction physics.  Rare-isotope facilities increasingly make such cross-method benchmarks possible.

\vspace*{0.06in}
\noindent
{\large\bf Short-range correlations and high-momentum nucleons}

Short-range correlations arise because realistic nuclear forces are strongly repulsive at very short distance and contain a powerful tensor component at intermediate range.  They produce pairs with large relative momentum
$
 \bk_{\rm rel}={(\bk_1-\bk_2)}/{2}
$
and comparatively small center-of-mass momentum
$
 \bK_{\rm cm}=\bk_1+\bk_2.
$
In the SRC domain one often finds approximate factorization of the two-body momentum distribution,
\begin{equation}
 n_{NN}(\bk_{\rm rel},\bK_{\rm cm})\approx C_{NN}\,
 n_{NN}^{\rm rel}(k_{\rm rel})n_{NN}^{\rm cm}(K_{\rm cm}),
\end{equation}
where $C_{NN}$ is related to the abundance of close-proximity pairs in a specified spin-isospin channel.

High-energy electron scattering established that two-nucleon SRCs dominate the high-momentum tail and that $np$ pairs greatly outnumber $pp$ pairs in a substantial momentum interval, consistent with tensor-force dominance (\cite{Subedi2008,Hen2017}).  The natural rare-isotope question is how this pair composition evolves as $N/Z$ becomes extreme.  A simple combinatorial argument would predict many more $nn$ than $np$ pairs in a neutron-rich nucleus, but tensor dynamics can cause minority protons to carry a disproportionately large share of high momentum.  This has implications for kinetic symmetry energy, neutrino response, and the interpretation of deeply bound nucleon removal.

Hadronic knockout can access related physics if measurements select high momentum transfer and reconstruct both members of the correlated pair.  A one-step process may strike one nucleon while the correlated partner recoils.  In an idealized factorized description,
\begin{equation}
 \dd\sigma_{A(p,pN N)}\sim K\,\sigma_{pN}\,
 F_{\rm FSI}\,n_{NN}(\bk_{\rm rel},\bK_{\rm cm})\dd\Phi,
\end{equation}
where $F_{\rm FSI}$ summarizes attenuation and rescattering.  The challenge is that hadronic probes experience much stronger initial- and final-state interactions than electrons.  Charge exchange, multistep scattering, and detector acceptance can mimic or distort pair correlations.  Consequently, SRC claims require kinematic cuts and reaction calculations designed specifically for the high-$Q^2$, large-$k_{\rm rel}$ regime.
Standard intermediate-energy heavy-ion two-nucleon removal is not primarily an SRC probe.  Its strong surface localization favors valence-space pair amplitudes and long-range correlations.  It is important to notice that ``two-nucleon correlation" in the eikonal pair-removal literature often refers to coherent shell-model spatial correlations rather than the universal high-momentum SRC phenomenology familiar from electron scattering (\cite{CiofiSimula1996}).  Both are genuine correlations, but they correspond to different regions of the two-body density matrix.

\vspace*{0.06in}
\noindent
{\large\bf Pairing, continuum coupling, halos, and skins}

Near the neutron drip line, pairing and continuum coupling become intertwined. In Hartree–Fock–Bogoliubov (HFB) theory, the quasiparticle Hamiltonian contains a particle-hole field, a pairing field, and a chemical potential $\lambda$  (\cite{RingSchuck1980,Barranco2001}). As $\lambda \rightarrow 0$, quasiparticle wave functions can extend into the continuum. Pairing may either enhance occupancy of spatially extended low-$l$ states or, through the so-called pairing anti-halo mechanism (\cite{PhysRevC.95.024304}), prevent an rms radius from diverging as strongly as an unpaired single-particle estimate would suggest.

One-nucleon knockout is highly sensitive to weakly bound $s$ and $p$ orbitals through narrow residue momentum distributions and large surface amplitudes.  Classic halo studies used this fact to identify extended valence configurations.  Yet a halo is itself a correlation phenomenon when more than one valence nucleon is involved.  In a Borromean two-neutron halo, neither neutron may be bound to the core individually, while the three-body system is bound.  The wave function depends on neutron-neutron correlations, core recoil, and continuum dynamics.

For a core$+n+n$ system (with coordinates ${\bf r}_c$,  ${\bf r}_1$,  and ${\bf r}_2$), Jacobi coordinates may be chosen as
$
 \bx=\br_1-\br_2,$ and
 $\by={(\br_1+\br_2)}/{2}-\br_c.
$
The opening-angle distribution and correlations between $x$ and $y$ distinguish dineutron-like configurations from more extended ``cigar'' geometries (\cite{BertulaniHussein2007}).  Knockout and breakup observables can constrain these patterns when sufficiently exclusive.  Removing one neutron from a two-neutron halo often leaves the core-neutron subsystem unbound, so invariant-mass spectroscopy of the decay products maps the continuum response.

Neutron skins provide a different form of spatial asymmetry.  A thick neutron skin increases the probability of peripheral neutron removal and modifies proton versus neutron reaction geometries.  Consequently, knockout systematics across isotopic chains contain information about radial distributions as well as shell occupancy.  This creates both an opportunity and a complication: a cross-section trend attributed to changing spectroscopic strength may partly reflect changing neutron and proton densities.  Reaction calculations should therefore use density distributions consistent with the same energy-density functional or many-body method used to predict the overlaps whenever possible (\cite{Aumann2013,Aumann2021}).

\vspace*{0.12in}
\noindent
{\large\bf Reaction-theory uncertainties and the meaning of factorization}

Every extraction of correlation information depends on a hierarchy of approximations.  The eikonal approximation assumes high energy and small deflection.  The sudden approximation assumes that intrinsic coordinates are frozen during the collision.  The optical-limit Glauber theory replaces complicated many-body scattering by folded densities and effective nucleon-nucleon profile functions.  DWIA assumes a dominant single hard collision embedded in distorted waves.  Factorization separates elementary scattering from nuclear structure.  Each approximation has a regime of validity, and unstable nuclei often push several limits simultaneously because separation energies may be tiny while recoil and continuum effects are large.  Corrections scale with deflection, potential strength, and inverse beam momentum.  Dynamical eikonal and continuum-discretized coupled-channel approaches can test the sudden and adiabatic limits (\cite{Bertulani2005,
PhysRevLett.95.082502}).  Medium corrections, finite-range effects, and Fermi motion alter the effective interaction at intermediate energy (\cite{Teixeira2022}).  These ingredients can matter at the 10-20\% level, comparable to structural effects one may wish to infer.

For quasifree scattering, uncertainties arise from optical potentials, effective $NN$ interactions, nonlocality, relativistic kinematics, off-shell prescriptions, and three-body dynamics (\cite{Aumann2013,Aumann2021}).  At lower energies, transfer-like and compound mechanisms can contaminate the impulse picture.  At high energies, Glauber multiple scattering and relativistic formulations become natural.  Benchmarking against stable nuclei with known electron-scattering spectral functions is therefore essential.

A useful formal viewpoint is to regard the measured quantity as
\begin{equation}
 \sigma=\mathrm{Tr}\left[\hat{\rho}^{(n)}\hat{K}_{\rm rxn}^{(n)}\right],
\end{equation}
where $\hat\rho^{(n)}$ is an $n$-body reduced density operator and $\hat K_{\rm rxn}^{(n)}$ is a reaction kernel.  One-nucleon removal primarily probes a one-body overlap, two-nucleon removal a transition two-body density, and SRC coincidence reactions a restricted high-momentum part of the two-body density.  This expression makes clear that no reaction observes the density matrix without a filter.

\vspace*{0.12in}
\noindent
{\large\bf Ab initio structure, effective interactions, and consistent operators}

The interpretation of knockout data is increasingly connected to ab initio calculations.  No-core shell model, coupled-cluster theory, self-consistent Green functions, in-medium similarity renormalization group, quantum Monte Carlo, and lattice-inspired methods now describe selected medium-mass nuclei with interactions rooted in chiral effective field theory (\cite{Alvioli2005,Epelbaum2009,Navratil2009,Bogner2010,Hagen2014,Carlson2015,Hergert2020,MachleidtEntem2011,Drischler2021,MachleidtSammarruca2024}).  These approaches predict energies, radii, one-body overlaps, spectral functions, and in some cases two-body overlap amplitudes.

Chiral effective field theory organizes the Hamiltonian as (\cite{BedaqueVanKolck2002,Epelbaum2009,Navratil2009,MachleidtEntem2011,MachleidtSammarruca2024})
\begin{equation}
 H=T+V_{NN}+V_{3N}+V_{4N}+\cdots,
\end{equation}
with interactions expanded in powers of $Q/\Lambda_\chi$.  Three-nucleon forces are essential for shell evolution, saturation, and the location of drip lines.  They therefore influence knockout observables indirectly through separation energies and configuration mixing and directly through correlated wave functions.

Similarity-renormalization transformations soften the Hamiltonian (\cite{BognerFurnstahlPerry2008}),
$
 H_s=U_sHU_s^\dagger,
$
but observables require consistently transformed operators,
$
 O_s=U_sOU_s^\dagger.
$
A spectroscopic factor computed from an untransformed annihilation operator can change with the resolution parameter $s$, while an exact cross section does not.  This is the formal basis for the statement that spectroscopic factors are scheme dependent.  Reaction theory should ultimately evolve toward consistent structure-plus-operator calculations rather than attaching a phenomenological reaction kernel to a wave function generated at an unrelated resolution.

The two-nucleon knockout studies using no-core shell-model overlaps for $^{12}$C were an early demonstration of this program (\cite{SimpsonNavratil2012}).  Different chiral Hamiltonians produced different two-nucleon amplitudes and hence different partial cross sections, especially to $T=0$ states.  Such sensitivity is scientifically valuable if reaction uncertainties are smaller than the structural spread.  It turns an exotic-beam reaction into a discriminator among microscopic Hamiltonians.
Uncertainty quantification should accompany this comparison.  A Bayesian framing is preferable to quoting a single reduction factor without an uncertainty budget (\cite{Furnstahl2015,PhysRevLett.122.232502}).  It can also identify which new measurements would most efficiently discriminate among correlation mechanisms.

\vspace*{0.12in}
\noindent
{\large\bf Experimental observables beyond inclusive cross sections}

Inclusive cross sections are experimentally efficient and remain valuable for surveying broad regions of the nuclear chart, but correlations are multidimensional.  The strongest future constraints will come from exclusive measurements.  $\gamma$-ray tagging identifies the bound final state of the residue, converting an inclusive removal yield into partial cross sections and state-specific momentum distributions (\cite{BertulaniDoornenbalObertelliUesaka2026}).  This separates configuration mixing among low-lying states and reduces dependence on theoretical branching assumptions.

For unbound residues, invariant-mass spectroscopy reconstructs the decay energy,
\begin{equation}
 E_{\rm rel}=\sqrt{\left(\sum_i E_i\right)^2-\left|\sum_i\bp_i c\right|^2}-\sum_i m_ic^2.
\end{equation}
The resulting continuum line shape can reveal resonances, virtual states, and final-state interactions.  In halo systems, correlations among decay fragments provide information about the initial three-body geometry, although the mapping is altered by continuum dynamics.

In $(p,2p)$ and $(p,pn)$ reactions, detection of both outgoing nucleons permits missing-momentum reconstruction (\cite{Panin2026}).  Coplanarity and energy-sharing conditions test the quasifree hypothesis.  The missing-energy spectrum separates hole states and continuum strength.  If an additional correlated nucleon is detected, one can construct relative and center-of-mass pair momenta and search for back-to-back SRC signatures.
Polarization observables offer further selectivity.  The elementary $NN$ amplitude contains central, spin-orbit, and spin-spin components.  Spin-transfer coefficients can therefore constrain the spin character of the removed configuration.  Although such measurements are difficult with low-intensity radioactive beams, future high-luminosity facilities may make them practical for selected cases (\cite{PhysRevC.107.054603,BertulaniDoornenbalObertelliUesaka2026}).

Residue transverse momentum and angular distributions probe reaction dynamics and can reveal deficiencies hidden in integrated cross sections.  A model may reproduce $\sigma$ while failing to reproduce the full distribution
$
{\dd^3\sigma}/{\dd p_x\dd p_y\dd p_z}.
$
Multidimensional likelihood analyses should therefore replace one-number comparisons whenever statistics permit.  Correlations among experimental bins must be retained, especially when acceptance corrections and unfolding are significant.

\vspace*{0.12in}
\noindent
{\large\bf Representative physics cases in unstable nuclei}

The oxygen chain illustrates how knockout can test shell evolution and correlation effects.  Oxygen has a well-known neutron drip line at $^{24}$O, while neighboring fluorine isotopes remain bound much farther in neutron number.  Three-nucleon forces and continuum coupling are central to this pattern.  Proton and neutron removal from oxygen isotopes therefore test both deeply bound and weakly bound overlaps in a system accessible to ab initio theory.  Quasifree proton-target measurements provide an especially useful benchmark for the separation-energy dependence of extracted strength (\cite{Atar2018}).

The carbon, nitrogen, and oxygen regions also contain classic halo and skin candidates.  Removal from low-$l$ orbitals produces narrow momentum distributions, while $\gamma$-ray tagging can distinguish core-excited components.  A wave function written schematically
\begin{equation}
 \ket{\Psi_A}=a\ket{\Phi_{\rm core}^{0^+}\otimes nlj}
 +\sum_\nu b_\nu\ket{\Phi_{\rm core}^{\nu}\otimes n'l'j'}+\cdots
\end{equation}
leads to partial knockout strengths proportional to $|a|^2$, $|b_\nu|^2$, and interference terms when channels are not orthogonal.  Measuring the distribution of strength among core states is therefore a direct test of particle-vibration coupling.

The magnesium region has been a proving ground for two-nucleon removal and the breakdown of traditional shell closures.  Around the ``island of inversion,'' intruder configurations from the $pf$ shell mix strongly with normal $sd$-shell configurations.  Pair-removal amplitudes can selectively populate final states according to this mixing.  Momentum distributions then provide an angular-momentum filter, making two-nucleon removal more than a means of reaching a rare isotope: it becomes a test of the correlated parent wave function (\cite{Simpson2009PRL}).

In neutron-rich calcium and beyond, the growing reach of rare-isotope facilities opens comparisons with coupled-cluster, IMSRG, and Green-function calculations (\cite{Otsuka2010,Hagen2012,Hergert2017,Soma2013,Ekstrom2015}).  Proton removal from neutron-rich systems is particularly interesting because the proton is often deeply bound relative to the neutron Fermi surface.  If tensor and SRC mechanisms preferentially deplete minority-proton quasiparticle strength, these nuclei should amplify the effect.  Yet they also amplify reaction-model challenges because the proton overlap is compact and surface sampling becomes severe.

Near the proton drip line, Coulomb barriers change the asymptotic behavior.  A weakly bound proton can remain spatially localized compared with a neutron of the same separation energy because of Coulomb confinement.  Proton removal therefore provides a complementary test of how geometry, separation energy, and correlations combine.  Mirror nuclei are particularly useful: differences between analogous neutron and proton removal can isolate Coulomb and isospin-breaking effects when the underlying shell structure is similar.

\vspace*{0.12in}
\noindent
{\large\bf A hierarchy of correlation observables}

It is helpful to classify what different knockout observables can realistically teach us.  At the first level are occupancies and configuration mixing.  Partial one-nucleon removal cross sections and $\gamma$-ray tagged final states constrain the distribution of single-particle strength.  The relevant correlation physics is primarily long-range: pairing, deformation, particle-vibration coupling, and cross-shell admixtures.

At the second level are radial correlations and weak binding.  Absolute cross sections, reaction probabilities, and momentum widths depend on the spatial extent of overlaps.  Halos, skins, and continuum coupling appear here.  These effects are not reducible to an occupancy; two wave functions with the same spectroscopic factor can produce different knockout cross sections if their radial overlaps differ.

At the third level are coherent pair correlations.  Two-nucleon removal partial cross sections and residue momentum distributions depend on two-nucleon amplitudes and their interference.  Small-angle pair localization at the projectile surface can enhance removal.  The $LS$ composition of the pair controls momentum shapes.  Isoscalar $np$ correlations can be compared with isovector like-particle pairing.

At the fourth level are short-range and tensor correlations.  These require high missing momentum, large momentum transfer, and preferably detection of both members of the pair.  The appropriate observables are pair relative momentum, center-of-mass momentum, angular correlations, and ratios such as
$
 R_{np/pp}(k_{\rm rel},K_{\rm cm})=
{\dd\sigma[np]}/{\dd\sigma[pp]},
$
with reaction corrections under control.  An inclusive two-nucleon removal cross section is generally insufficient for this purpose.

At the fifth level are genuinely many-body correlation patterns involving clusters or several nucleons.  Quasifree removal of $d$, $t$, $^3$He, or $\alpha$ clusters and multi-nucleon coincidence reactions can test preformation and higher-order reduced densities.  Recent reaction-theory developments explicitly emphasize the need to include two- and multi-nucleon correlations for such processes (\cite{10.1093/ptep/ptaf119}).  This frontier connects shell-model correlations with emergent clustering.

\vspace*{0.12in}
\noindent
{\large\bf Conclusions}

Knockout reactions occupy a special position in the study of unstable nuclei because they combine strong selectivity to single-particle motion with access to nuclei that cannot be used as stationary targets.  Their simplest interpretation is spectroscopic: remove a nucleon, identify the residue state, and infer the orbital from the momentum distribution.  Their deeper significance is many-body: the measured strength and its fragmentation are controlled by correlations, the radial overlap reflects binding and continuum coupling, and pair-removal observables contain interference information from the two-body wave function.

The eikonal theory of fast removal on light targets provides a transparent geometric picture.  Stripping and diffraction act at the projectile surface, and the cross section is a weighted norm of the overlap.  This makes the method efficient and often robust for orbital assignments, but it also means that extracted absolute strengths depend on radial geometry and reaction transparency.  The reduction factor $R_s$ is therefore an informative diagnostic, not a model-independent measurement of correlation strength.

Proton-induced quasifree scattering supplies an essential complement.  $(p,2p)$ and $(p,pn)$ reactions connect naturally to spectral functions and, with complete kinematics, can map missing energy and momentum.  They provide a different radial filter and a different set of reaction uncertainties.  Cross-comparison with heavy-ion removal and transfer is one of the strongest available tests of spectroscopic quenching in asymmetric nuclei.

Two-nucleon removal opens the transition from one-body to pair correlations.  Its sensitivity arises because the reaction demands simultaneous geometric access to two nucleons while the residue survives.  Coherent two-nucleon amplitudes determine the surface pair density, and their $LS$ composition shapes the residue momentum distribution.  The method has demonstrated sensitivity to isoscalar $np$ correlations and to microscopic two-nucleon overlaps.  It should not, however, be conflated automatically with short-range-correlation measurements; valence pair correlations and high-momentum SRCs occupy different sectors of the two-body density.

The broad lesson is methodological.  ``Nucleon correlations'' are not a single correction to the shell model.  They comprise long-range collective mixing, pairing, continuum coupling, tensor correlations, short-range repulsion, and emergent clustering.  Different knockout observables project different combinations of these mechanisms.  The path toward quantitative understanding is therefore to combine exclusive experiments, microscopic structure theory, controlled reaction models, and uncertainty quantification.  With that synthesis, unstable nuclei become more than exotic endpoints of the chart of nuclides: they become tunable many-body systems in which the spatial, spin, isospin, and momentum structure of nuclear correlations can be tested under extreme conditions.

\bibliographystyle{Harvard}
\bibliography{references}

@book{BohrMottelson1969,
  author    = {Bohr, A. and Mottelson, B. R.},
  title     = {Nuclear Structure, Vol. I: Single-Particle Motion},
  volume    = {1},
  publisher = {W. A. Benjamin},
  address   = {New York},
  year      = {1969}
}

@book{BohrMottelson1975,
  author    = {Bohr, A. and Mottelson, B. R.},
  title     = {Nuclear Structure, Vol. II: Nuclear Deformations},
  volume    = {2},
  publisher = {W. A. Benjamin},
  address   = {Reading, Massachusetts},
  year      = {1975}
}

@article{HansenTostevin2003,
  author  = {Hansen, P. G. and Tostevin, J. A.},
  title   = {Direct Reactions with Exotic Nuclei},
  journal = {Annual Review of Nuclear and Particle Science},
  volume  = {53},
  pages   = {219--261},
  year    = {2003},
  doi     = {10.1146/annurev.nucl.53.041002.110406}
}

@article{PhysRevC.95.024304,
  title = {New concept for the pairing anti-halo effect as a localized wave packet of quasiparticles},
  author = {Hagino, K. and Sagawa, H.},
  journal = {Phys. Rev. C},
  volume = {95},
  issue = {2},
  pages = {024304},
  numpages = {5},
  year = {2017},
  month = {Feb},
  publisher = {American Physical Society},
  doi = {10.1103/PhysRevC.95.024304}
}

@article{BertulaniHussein2007,
  author  = {Bertulani, C. A. and Hussein, M. S.},
  title   = {Geometry of Borromean Halo Nuclei},
  journal = {Physical Review C},
  volume  = {76},
  pages   = {051602},
  year    = {2007},
  doi     = {10.1103/PhysRevC.76.051602}
}

@article{Bertulani2005,
  author  = {Bertulani, C. A.},
  title   = {Relativistic Continuum-Continuum Coupling in the
             Dissociation of Halo Nuclei},
  journal = {Physical Review Letters},
  volume  = {94},
  pages   = {072701},
  year    = {2005},
  doi     = {10.1103/PhysRevLett.94.072701}
}

@article{MachleidtSammarruca2024,
  author  = {Machleidt, R. and Sammarruca, F.},
  title   = {Recent Advances in Chiral {EFT} Based Nuclear Forces
             and Their Applications},
  journal = {Progress in Particle and Nuclear Physics},
  volume  = {137},
  pages   = {104117},
  year    = {2024},
  doi     = {10.1016/j.ppnp.2024.104117}
}

@article{BedaqueVanKolck2002,
  author  = {Bedaque, Paulo F. and van Kolck, Ubirajara},
  title   = {Effective Field Theory for Few-Nucleon Systems},
  journal = {Annual Review of Nuclear and Particle Science},
  volume  = {52},
  pages   = {339--396},
  year    = {2002},
  doi     = {10.1146/annurev.nucl.52.050102.090637}
}

@article{Teixeira2022,
  author  = {Teixeira, E. A. and Aumann, T. and Bertulani, C. A. and Carlson, B. V.},
  title   = {Nuclear Fragmentation Reactions as a Probe of Neutron Skins in Nuclei},
  journal = {The European Physical Journal A},
  volume  = {58},
  pages   = {205},
  year    = {2022},
  doi     = {10.1140/epja/s10050-022-00849-w}
}

@article{Navratil2009,
  author  = {Navr{\'a}til, P. and Quaglioni, S. and Stetcu, I. and Barrett, B. R.},
  title   = {Recent Developments in No-Core Shell-Model Calculations},
  journal = {Journal of Physics G: Nuclear and Particle Physics},
  volume  = {36},
  number  = {8},
  pages   = {083101},
  year    = {2009},
  doi     = {10.1088/0954-3899/36/8/083101}
}

@article{BognerFurnstahlPerry2008,
  author  = {Bogner, S. K. and Furnstahl, R. J. and Perry, R. J.},
  title   = {Three-Body Forces Produced by a Similarity Renormalization
             Group Transformation in a Simple Model},
  journal = {Annals of Physics},
  volume  = {323},
  number  = {6},
  pages   = {1478--1501},
  year    = {2008},
  doi     = {10.1016/j.aop.2007.09.001}
}

@article{PhysRevLett.95.082502,
  title = {Collisions of Halo Nuclei within a Dynamical Eikonal Approximation},
  author = {Baye, D. and Capel, P. and Goldstein, G.},
  journal = {Phys. Rev. Lett.},
  volume = {95},
  issue = {8},
  pages = {082502},
  numpages = {4},
  year = {2005},
  month = {Aug},
  publisher = {American Physical Society},
  doi = {10.1103/PhysRevLett.95.082502}
}

@article{AlKhaliliNunes2003,
  author  = {Al-Khalili, J. S. and Nunes, F. M.},
  title   = {Reaction Models to Probe the Structure of Light Exotic Nuclei},
  journal = {Journal of Physics G: Nuclear and Particle Physics},
  volume  = {29},
  pages   = {R89--R132},
  year    = {2003},
  doi     = {10.1088/0954-3899/29/11/R01}
}

@article{HansenJensenJonson1995,
  author  = {Hansen, P. G. and Jensen, A. S. and Jonson, B.},
  title   = {Nuclear Halos},
  journal = {Annual Review of Nuclear and Particle Science},
  volume  = {45},
  pages   = {591--634},
  year    = {1995},
  doi     = {10.1146/annurev.ns.45.120195.003111}
}

@article{JensenRiisagerFedorovGarrido2004,
  author  = {Jensen, A. S. and Riisager, K. and Fedorov, D. V. and Garrido, E.},
  title   = {Structure and Reactions of Quantum Halos},
  journal = {Reviews of Modern Physics},
  volume  = {76},
  pages   = {215--261},
  year    = {2004},
  doi     = {10.1103/RevModPhys.76.215}
}

@article{BaranColonnaGrecoDiToro2005,
  author  = {Baran, V. and Colonna, M. and Greco, V. and Di Toro, M.},
  title   = {Reaction Dynamics with Exotic Nuclei},
  journal = {Physics Reports},
  volume  = {410},
  pages   = {335--466},
  year    = {2005},
  doi     = {10.1016/j.physrep.2004.12.004}
}

@article{BertulaniCantoHussein1993,
  author  = {Bertulani, C. A. and Canto, L. F. and Hussein, M. S.},
  title   = {The Structure and Reactions of Neutron-Rich Nuclei},
  journal = {Physics Reports},
  volume  = {226},
  pages   = {281--376},
  year    = {1993},
  doi     = {10.1016/0370-1573(93)90128-Z}
}

@article{BertulaniGade2010,
  author  = {Bertulani, C. A. and Gade, A.},
  title   = {Nuclear Astrophysics with Radioactive Beams},
  journal = {Physics Reports},
  volume  = {485},
  pages   = {195--259},
  year    = {2010},
  doi     = {10.1016/j.physrep.2009.09.002}
}

@article{BertulaniBonaccorso2022,
  author  = {Bertulani, C. A. and Bonaccorso, A.},
  title   = {Direct Nuclear Reactions with Rare Isotopes},
  journal = {Handbook of Nuclear Physics, Springer Nature},
  Volume = { },
  pages   = {1-35},
  year    = {2020},
  doi     = {10.1007/978-981-15-8818-1_3-1}
}

@book{Bertulani2007,
  author    = {Bertulani, C. A.},
  title     = {Nuclear Physics in a Nutshell},
  publisher = {Princeton University Press},
  address   = {Princeton, New Jersey},
  year      = {2007},
  isbn      = {978-0-691-12505-3},
  doi       = {10.1515/9781400839322}
}

@book{ObertelliSagawa2021,
  author    = {Obertelli, A. and Sagawa, H.},
  title     = {Modern Nuclear Physics: From Fundamentals to Frontiers},
  series    = {UNITEXT for Physics},
  publisher = {Springer},
  address   = {Singapore},
  year      = {2021},
  doi       = {10.1007/978-981-16-2289-2},
  isbn      = {978-981-16-2289-2}
}

@article{Aumann2021,
  author  = {Aumann, T. and others},
  title   = {Quenching of Single-Particle Strength from Direct Reactions
             with Stable and Rare-Isotope Beams},
  journal = {Progress in Particle and Nuclear Physics},
  volume  = {118},
  pages   = {103847},
  year    = {2021},
  doi     = {10.1016/j.ppnp.2021.103847}
}

@book{Hagino2013,
  author    = {Hagino, K. and Tanihata, I. and Sagawa, H.},
  title     = {Exotic Nuclei Far from the Stability Line},
  booktitle = {100 Years of Subatomic Physics},
  editor    = {Henley, E. M. and Ellis, S. D.},
  pages     = {231--272},
  publisher = {World Scientific},
  address   = {Singapore},
  year      = {2013},
  doi       = {10.1142/9789814425810_0009}
}

@article{Alamanos2022,
  author  = {Alamanos, N. and others},
  title   = {Editorial: Re-writing Nuclear Physics Textbooks:
             Recent Advances in Nuclear Physics Applications},
  journal = {The European Physical Journal Plus},
  volume  = {137},
  pages   = {350},
  year    = {2022},
  doi     = {10.1140/epjp/s13360-022-02552-7}
}

@article{HansenJonson1987,
  author  = {Hansen, P. G. and Jonson, B.},
  title   = {The Neutron Halo of Extremely Neutron-Rich Nuclei},
  journal = {Europhysics Letters},
  volume  = {4},
  number  = {4},
  pages   = {409--414},
  year    = {1987},
  doi     = {10.1209/0295-5075/4/4/005}
}

@article{Tostevin2001,
  author  = {Tostevin, J. A.},
  title   = {Single-Nucleon Knockout Reactions at Fragmentation
             Beam Energies},
  journal = {Nuclear Physics A},
  volume  = {682},
  pages   = {320c--331c},
  year    = {2001},
  doi     = {10.1016/S0375-9474(00)00656-4}
}

@article{GadeTostevin2008,
  author  = {Gade, A. and Tostevin, J. A.},
  title   = {Knockout Reactions},
  journal = {Nuclear Physics News},
  volume  = {20},
  number  = {1},
  pages   = {11--16},
  year    = {2010},
  doi     = {10.1080/10506890903178889}
}

@article{Aumann2013,
  author  = {Aumann, T. and Bertulani, C. A. and Ryckebusch, J.},
  title   = {Quasifree $(p,2p)$ and $(p,pn)$ Reactions with Unstable Nuclei},
  journal = {Physical Review C},
  volume  = {88},
  number  = {6},
  pages   = {064610},
  year    = {2013},
  doi     = {10.1103/PhysRevC.88.064610}
}

@article{DickhoffBarbieri2004,
  author  = {Dickhoff, W. H. and Barbieri, C.},
  title   = {Self-Consistent Green's Function Method for Nuclei
             and Nuclear Matter},
  journal = {Progress in Particle and Nuclear Physics},
  volume  = {52},
  number  = {2},
  pages   = {377--496},
  year    = {2004},
  doi     = {10.1016/j.ppnp.2004.02.038}
}

@article{Soma2020,
  author  = {Som{\`a}, V.},
  title   = {Self-Consistent Green's Function Theory for Atomic Nuclei},
  journal = {Frontiers in Physics},
  volume  = {8},
  pages   = {340},
  year    = {2020},
  doi     = {10.3389/fphy.2020.00340}
}

@book{BarbieriCarbone2017,
  author    = {Barbieri, C. and Carbone, A.},
  title     = {Self-Consistent Green's Function Approaches},
  booktitle = {An Advanced Course in Computational Nuclear Physics},
  editor    = {Hjorth-Jensen, M. and Lombardo, M. P. and van Kolck, U.},
  series    = {Lecture Notes in Physics},
  volume    = {936},
  pages     = {571--644},
  publisher = {Springer},
  year      = {2017},
  doi       = {10.1007/978-3-319-53336-0_11}
}

@article{BarbieriDickhoff2009,
  author  = {Barbieri, C. and Dickhoff, W. H.},
  title   = {Spectroscopic Factors in {$^{16}$O} and Nucleon Asymmetry},
  journal = {International Journal of Modern Physics A},
  volume  = {24},
  number  = {11},
  pages   = {2060--2068},
  year    = {2009},
  doi     = {10.1142/S0217751X09045625}
}

@article{Duguet2015,
   author  = {Duguet, T. and Hergert, H. and Holt, J. D. and Som{\`a}, V.},
  title   = {Nonobservable Nature of the Nuclear Shell Structure:
             Meaning, Illustrations, and Consequences},
  journal = {Physical Review C},
  volume  = {92},
  number  = {3},
  pages   = {034313},
  year    = {2015},
  doi     = {10.1103/PhysRevC.92.034313}
}

@article{Subedi2008,
  author  = {Subedi, R. and others},
  title   = {Probing Cold Dense Nuclear Matter},
  journal = {Science},
  volume  = {320},
  number  = {5882},
  pages   = {1476--1478},
  year    = {2008},
  doi     = {10.1126/science.1156675}
}

@book{RingSchuck1980,
  author    = {Ring, Peter and Schuck, Peter},
  title     = {The Nuclear Many-Body Problem},
  publisher = {Springer-Verlag},
  address   = {Berlin, Heidelberg},
  year      = {1980},
  series    = {Theoretical and Mathematical Physics}
}

@book{Glauber1959,
  author    = {Glauber, R. J.},
  title     = {High-Energy Collision Theory},
  booktitle = {Lectures in Theoretical Physics},
  editor    = {Brittin, W. E. and Dunham, L. G.},
  volume    = {1},
  pages     = {315--414},
  publisher = {Interscience Publishers},
  address   = {New York},
  year      = {1959}
}

@article{BertulaniDeConti2010,
  author  = {Bertulani, C. A. and De Conti, C.},
  title   = {Pauli Blocking and Medium Effects in Nucleon Knockout Reactions},
  journal = {Physical Review C},
  volume  = {81},
  number  = {6},
  pages   = {064603},
  year    = {2010},
  doi     = {10.1103/PhysRevC.81.064603}
}

@article{Gade2008,
  author  = {Gade, A. and others},
  title   = {Reduction of Spectroscopic Strength: Weakly-Bound and
             Strongly-Bound Single-Particle States Studied Using
             One-Nucleon Knockout Reactions},
  journal = {Physical Review C},
  volume  = {77},
  number  = {4},
  pages   = {044306},
  year    = {2008},
  doi     = {10.1103/PhysRevC.77.044306}
}

@article{JacobGade2014,
  author  = {Gade, A. and Glasmacher, T.},
  title   = {In-Beam Nuclear Spectroscopy of Bound States
             with Fast Exotic Ion Beams},
  journal = {Progress in Particle and Nuclear Physics},
  volume  = {60},
  number  = {1},
  pages   = {161--224},
  year    = {2008},
  doi     = {10.1016/j.ppnp.2007.08.001}
}

@book{Moro2019,
  author    = {Moro, A. M.},
  title     = {Models for Nuclear Reactions with Weakly Bound Systems},
  booktitle = {Nuclear Physics with Stable and Radioactive Ion Beams},
  series    = {Proceedings of the International School of Physics
               ``Enrico Fermi''},
  volume    = {201},
  pages     = {129--207},
  publisher = {IOS Press},
  year      = {2019},
  doi       = {10.3254/978-1-61499-957-7-129}
}

@article{WakasaOgataNoro2017,
  author  = {Wakasa, T. and Ogata, K. and Noro, T.},
  title   = {Proton-Induced Knockout Reactions with Polarized
             and Unpolarized Beams},
  journal = {Progress in Particle and Nuclear Physics},
  volume  = {96},
  pages   = {32--87},
  year    = {2017},
  doi     = {10.1016/j.ppnp.2017.06.002}
}

@article{TostevinGade2014,
  author  = {Tostevin, J. A. and Gade, A.},
  title   = {Systematics of Intermediate-Energy Single-Nucleon
             Removal Cross Sections},
  journal = {Physical Review C},
  volume  = {90},
  number  = {5},
  pages   = {057602},
  year    = {2014},
  doi     = {10.1103/PhysRevC.90.057602}
}

@article{Kay2013,
  author  = {Kay, B. P. and Schiffer, J. P. and Freeman, S. J.},
  title   = {Quenching of Cross Sections in Nucleon Transfer Reactions},
  journal = {Physical Review Letters},
  volume  = {111},
  number  = {4},
  pages   = {042502},
  year    = {2013},
  doi     = {10.1103/PhysRevLett.111.042502}
}

@article{Atar2018,
  author = {Atar, L. and  others},
  title = {{Quasifree $(p,2p)$ Reactions on Oxygen Isotopes: Observation of Isospin Independence of the Reduced Single-Particle Strength}},
  year = {2018},
  journal = {{Physical Review Letters}},
  volume = {120},
  pages = {{052501}},
    doi     = {10.1103/PhysRevLett.120.052501}
}

@article{GomezRamosMoro2018,
  author  = {G{\'o}mez-Ramos, M. and Moro, A. M.},
  title   = {Binding-Energy Independence of Reduced Spectroscopic
             Strengths Derived from $(p,2p)$ and $(p,pn)$ Reactions
             with Nitrogen and Oxygen Isotopes},
  journal = {Physics Letters B},
  volume  = {785},
  pages   = {511--516},
  year    = {2018},
  doi     = {10.1016/j.physletb.2018.08.058}
}

@article{Hencken1996,
  author  = {Hencken, K. and Bertsch, G. F. and Esbensen, H.},
  title   = {Breakup Reactions of the Halo Nuclei
             {$^{11}$Be} and {$^{8}$B}},
  journal = {Physical Review C},
  volume  = {54},
  number  = {6},
  pages   = {3043--3050},
  year    = {1996},
  doi     = {10.1103/PhysRevC.54.3043}
}

@article{BertulaniMcVoy1992,
  author  = {Bertulani, C. A. and McVoy, K. W.},
  title   = {Momentum Distributions in Reactions with Radioactive Beams},
  journal = {Physical Review C},
  volume  = {46},
  number  = {6},
  pages   = {2638--2641},
  year    = {1992},
  doi     = {10.1103/PhysRevC.46.2638}
}

@article{BertulaniHansen2004,
  author  = {Bertulani, C. A. and Hansen, P. G.},
  title   = {Momentum Distributions in Stripping Reactions
             of Radioactive Projectiles at Intermediate Energies},
  journal = {Physical Review C},
  volume  = {70},
  number  = {3},
  pages   = {034609},
  year    = {2004},
  doi     = {10.1103/PhysRevC.70.034609}
}

@article{SimpsonTostevin2010,
  author  = {Simpson, E. C. and Tostevin, J. A.},
  title   = {Correlations Probed in Direct Two-Nucleon Removal Reactions},
  journal = {Physical Review C},
  volume  = {82},
  number  = {4},
  pages   = {044616},
  year    = {2010},
  doi     = {10.1103/PhysRevC.82.044616}
}

@article{SimpsonTostevin2011,
  author  = {Simpson, E. C. and Tostevin, J. A.},
  title   = {Two-Nucleon Correlation Effects in Knockout Reactions
             from {$^{12}$C}},
  journal = {Physical Review C},
  volume  = {83},
  number  = {1},
  pages   = {014605},
  year    = {2011},
  doi     = {10.1103/PhysRevC.83.014605}
}

@article{Longfellow2020,
  author = {Longfellow, B.  and others},
  title = {{Two-Neutron Knockout as a Probe of the Composition of States in $^{22}$Mg, $^{23}$Al, and $^{24}$Si}},
  year = {2020},
  journal = {{Physical Review C}},
  volume = {101},
  pages = {{031303(R)}},
  doi     = {10.1103/PhysRevC.101.031303}
}

@article{SimpsonNavratil2012,
  author = {Simpson, E. C. and Navr\'atil, P. and Roth, R. and Tostevin, J. A.},
  title = {{Microscopic Two-Nucleon Overlaps and Knockout Reactions from $^{12}$C}},
  year = {2012},
  journal = {{Physical Review C}},
  volume = {86},
  pages = {{054609}},
    doi     = {10.1103/PhysRevC.86.054609}
}

@article{Simpson2009PRL,
  author  = {Simpson, E. C. and
             Tostevin, J. A. and
             Bazin, D. and
             Brown, B. A. and
             Gade, A.},
  title   = {Two-Nucleon Knockout Spectroscopy at the Limits
             of Nuclear Stability},
  journal = {Physical Review Letters},
  volume  = {102},
  number  = {13},
  pages   = {132502},
  year    = {2009},
  doi     = {10.1103/PhysRevLett.102.132502}
}

@article{Simpson2009,
  author  = {Simpson, E. C. and
             Tostevin, J. A. and
             Bazin, D. and
             Gade, A.},
  title   = {Longitudinal Momentum Distributions of the Reaction
             Residues Following Fast Two-Nucleon Knockout Reactions},
  journal = {Physical Review C},
  volume  = {79},
  number  = {6},
  pages   = {064621},
  year    = {2009},
  doi     = {10.1103/PhysRevC.79.064621}
}

@article{FurnstahlHammer2001,
  author  = {Furnstahl, R. J. and Hammer, H.-W.},
  title   = {Are Occupation Numbers Observable?},
  journal = {Physics Letters B},
  volume  = {531},
  number  = {3--4},
  pages   = {203--208},
  year    = {2002},
  doi     = {10.1016/S0370-2693(01)01504-0}
}

@article{Panin2026,
  author  = {Panin, V. and others},
  title   = {Probing {QFS} Mechanism with the
             {$^{12}$C$(p,2p)^{11}$B} Reaction in Inverse Kinematics},
  journal = {Physics Letters B},
  volume  = {879},
  pages   = {140595},
  year    = {2026},
  doi     = {10.1016/j.physletb.2026.140595}
}

@article{PhysRevC.107.054603,
  title = {Effective polarization in proton-induced $\ensuremath{\alpha}$ knockout reactions},
  author = {Edagawa, Tomoatsu and Yoshida, Kazuki and Chazono, Yoshiki and Ogata, Kazuyuki},
  journal = {Phys. Rev. C},
  volume = {107},
  issue = {5},
  pages = {054603},
  numpages = {7},
  year = {2023},
  month = {May},
  publisher = {American Physical Society},
  doi = {10.1103/PhysRevC.107.054603}
}

@book{AlKhaliliTostevin2002,
  author    = {Al-Khalili, J. S. and Tostevin, J. A.},
  title     = {Few-Body Models of Nuclear Reactions},
  booktitle = {The Euroschool Lectures on Physics with Exotic Beams, Al-Khalili, J. S., and Roeckl, E. (eds.), Springer},
  pages     = {1373--1392},
  year      = {2002},
  publisher = {Springer},
  doi      = {10.1016/B978-012613760-6/50074-7}
}

@article{BertulaniDoornenbalObertelliUesaka2026,
  author  = {Bertulani, C. A. and
             Doornenbal, P. and
             Obertelli, A. and
             Uesaka, T.},
  title   = {Direct Reactions and Spectroscopy with Hydrogen Targets
             at the {RIBF}},
  journal = {Progress of Theoretical and Experimental Physics},
  volume  = {2026},
  number  = {4},
  pages   = {04A106},
  year    = {2026},
  doi     = {10.1093/ptep/ptaf091}
}

@article{PhysRevLett.122.232502,
  title = {Direct Comparison between Bayesian and Frequentist Uncertainty Quantification for Nuclear Reactions},
  author = {King, G. B. and Lovell, A. E. and Neufcourt, L. and Nunes, F. M.},
  journal = {Phys. Rev. Lett.},
  volume = {122},
  issue = {23},
  pages = {232502},
  numpages = {5},
  year = {2019},
  month = {Jun},
  publisher = {American Physical Society},
  doi = {10.1103/PhysRevLett.122.232502},
}

@article{PhysRevC.105.024613,
  title = {Nuclear spectroscopy with heavy ion nucleon knockout and (p,2p) reactions},
  author = {Li, Jianguo and Bertulani, Carlos A. and Xu, Furong},
  journal = {Phys. Rev. C},
  volume = {105},
  issue = {2},
  pages = {024613},
  numpages = {9},
  year = {2022},
  month = {Feb},
  publisher = {American Physical Society},
  doi = {10.1103/PhysRevC.105.024613},
}

@article{LiPeiPang2026,
  author  = {Li, Shichang and Pei, Junchen and Pang, Danyang},
  title   = {Explanation for Spectroscopic Factor Quenching
             in Knockout Reactions},
  journal = {Science China Physics, Mechanics \& Astronomy},
  volume  = {69},
  number  = {4},
  pages   = {242011},
  year    = {2026},
  doi     = {10.1007/s11433-025-2879-0}
}

@article{PhysRevC.100.064604,
  title = {Toward a reliable description of $(p,pN)$ reactions in the distorted-wave impulse approximation},
  author = {Phuc, Nguyen Tri Toan and Yoshida, Kazuki and Ogata, Kazuyuki},
  journal = {Phys. Rev. C},
  volume = {100},
  issue = {6},
  pages = {064604},
  numpages = {8},
  year = {2019},
  month = {Dec},
  publisher = {American Physical Society},
  doi = {10.1103/PhysRevC.100.064604},
}

@article{HEBBORN2024138413,
title = {Sensitivity of one-neutron knockout observables of loosely- to more deeply-bound nuclei},
journal = {Physics Letters B},
volume = {848},
pages = {138413},
year = {2024},
issn = {0370-2693},
doi = {10.1016/j.physletb.2023.138413},
author = {C. Hebborn and P. Capel},
}

@article{PASCHALIS2020135110,
title = {Nucleon-nucleon correlations and the single-particle strength in atomic nuclei},
journal = {Physics Letters B},
volume = {800},
pages = {135110},
year = {2020},
issn = {0370-2693},
doi = {10.1016/j.physletb.2019.135110},
author = {S. Paschalis and M. Petri and A.O. Macchiavelli and O. Hen and E. Piasetzky},
}

@article{BERTULANI2023138250,
title = {Core destruction in knockout reactions},
journal = {Physics Letters B},
volume = {846},
pages = {138250},
year = {2023},
issn = {0370-2693},
doi = {10.1016/j.physletb.2023.138250},
author = {C.A. Bertulani}
}

@article{PhysRevC.93.044607,
  title = {Experimental study of the knockout reaction mechanism using $^{14}\text{O}$ at 60 MeV/nucleon},
  author = {Sun, Y. L. and others},
  journal = {Phys. Rev. C},
  volume = {93},
  issue = {4},
  pages = {044607},
  numpages = {8},
  year = {2016},
  month = {Apr},
  publisher = {American Physical Society},
  doi = {10.1103/PhysRevC.93.044607},
}

@article{GOMEZRAMOS2023138284,
title = {Isospin dependence in single-nucleon removal cross sections explained through valence-core destruction effects},
journal = {Physics Letters B},
volume = {847},
pages = {138284},
year = {2023},
issn = {0370-2693},
doi = {10.1016/j.physletb.2023.138284},
author = {M. G\'omez-Ramos and J. G\'omez-Camacho and A.M. Moro}
}

@article{PhysRevC.73.044608,
  title = {Reduced neutron spectroscopic factors when using potential geometries constrained by {H}artree-{F}ock calculations},
  author = {Lee, Jenny and others},
  journal = {Phys. Rev. C},
  volume = {73},
  issue = {4},
  pages = {044608},
  numpages = {6},
  year = {2006},
  month = {Apr},
  publisher = {American Physical Society},
  doi = {10.1103/PhysRevC.73.044608},
}

@article{PhysRevLett.131.212503,
  title = {New Perspectives on Spectroscopic Factor Quenching from Reactions},
  author = {Hebborn, C. and Nunes, F. M. and Lovell, A. E.},
  journal = {Phys. Rev. Lett.},
  volume = {131},
  issue = {21},
  pages = {212503},
  numpages = {6},
  year = {2023},
  month = {Nov},
  publisher = {American Physical Society},
  doi = {10.1103/PhysRevLett.131.212503},
}

@article{XIE2023137800,
title = {Investigation of spectroscopic factors of deeply-bound nucleons in drip-line nuclei with the Gamow shell model},
journal = {Physics Letters B},
volume = {839},
pages = {137800},
year = {2023},
issn = {0370-2693},
doi = {10.1016/j.physletb.2023.137800},
author = {M.R. Xie and others},
}

@article{RODRIGUEZSANCHEZ2024138559,
title = {Short-range correlations in dynamical intranuclear cascade models for describing nucleon knockout reactions},
journal = {Physics Letters B},
volume = {851},
pages = {138559},
year = {2024},
issn = {0370-2693},
doi = {10.1016/j.physletb.2024.138559},
author = {J.L. Rodr{\'i}guez-S{\'a}nchez and J. Cugnon and J.C. David and J. Hirtz},
}

@article{Holl2019,
  author  = {Holl, M. and others},
  title   = {Quasi-Free Neutron and Proton Knockout Reactions
             from Light Nuclei in a Wide Neutron-to-Proton
             Asymmetry Range},
  journal = {Physics Letters B},
  volume  = {795},
  pages   = {682--688},
  year    = {2019},
  doi     = {10.1016/j.physletb.2019.06.069}
}

@article{PaninAumannBertulani2021,
  author  = {Panin, V. and Aumann, T. and Bertulani, C. A.},
  title   = {Quasi-Free Scattering in Inverse Kinematics
             as a Tool to Unveil the Structure of Nuclei},
  journal = {European Physical Journal A},
  volume  = {57},
  number  = {3},
  pages   = {103},
  year    = {2021},
  doi     = {10.1140/epja/s10050-021-00416-9}
}

@article{Capel2022,
  author  = {Capel, Pierre},
  title   = {Combining {Halo-EFT} Descriptions of Nuclei
             and Precise Models of Nuclear Reactions},
  journal = {Few-Body Systems},
  volume  = {63},
  number  = {1},
  pages   = {14},
  year    = {2022},
  doi     = {10.1007/s00601-021-01718-w}
}

@article{PhysRevC.98.034610,
  title = {Dissecting reaction calculations using halo effective field theory and ab initio input},
  author = {Capel, P. and Phillips, D. R. and Hammer, H.-W.},
  journal = {Phys. Rev. C},
  volume = {98},
  issue = {3},
  pages = {034610},
  numpages = {17},
  year = {2018},
  month = {Sep},
  publisher = {American Physical Society},
  doi = {10.1103/PhysRevC.98.034610},
}

@article{Panin2019,
  author  = {Panin, V. and  others},
  title   = {Quasi-Free Proton Knockout from {$^{12}$C} on Carbon
             Target at 398 {MeV/u}},
  journal = {Physics Letters B},
  volume  = {797},
  pages   = {134802},
  year    = {2019},
  doi     = {10.1016/j.physletb.2019.134802}
}

@article{Panin2016,
  author  = {Panin, V. and others},
  title   = {Exclusive Measurements of Quasi-Free Proton Scattering
             Reactions in Inverse and Complete Kinematics},
  journal = {Physics Letters B},
  volume  = {753},
  pages   = {204--210},
  year    = {2016},
  doi     = {10.1016/j.physletb.2015.11.082}
}

@book{Bertulani2003,
  author    = {Bertulani, C. A. and Danielewicz, P.},
  title     = {Introduction to Nuclear Reactions},
  edition   = {2},
  publisher = {CRC Press},
  address   = {Boca Raton},
  year      = {2021},
  doi       = {10.1201/9780429331060},
  isbn      = {978-0-367-35362-9}
}

@article{FrankfurtStrikman1988,
  author  = {Frankfurt, L. L. and Strikman, M. I.},
  title   = {Hard Nuclear Processes and Microscopic Nuclear Structure},
  journal = {Physics Reports},
  volume  = {160},
  number  = {5--6},
  pages   = {235--427},
  year    = {1988},
  doi     = {10.1016/0370-1573(88)90179-2}
}

@article{Arrington2012,
  author  = {Arrington, J. and Higinbotham, D. W. and Rosner, G. and Sargsian, M.},
  title   = {Hard Probes of Short-Range Nucleon-Nucleon Correlations},
  journal = {Progress in Particle and Nuclear Physics},
  volume  = {67},
  number  = {4},
  pages   = {898--938},
  year    = {2012},
  doi     = {10.1016/j.ppnp.2012.04.002}
}

@article{Hen2017,
  author  = {Hen, O. and Miller, G. A. and Piasetzky, E. and Weinstein, L. B.},
  title   = {Nucleon-Nucleon Correlations, Short-Lived Excitations,
             and the Quarks Within},
  journal = {Reviews of Modern Physics},
  volume  = {89},
  number  = {4},
  pages   = {045002},
  year    = {2017},
  doi     = {10.1103/RevModPhys.89.045002}
}

@book{Dickhoff2017,
  author = {Dickhoff, W. H. and Van Neck, D.},
  title = {{Many-Body Theory Exposed! Propagator Description of Quantum Mechanics in Many-Body Systems}},
  year = {2008},
   doi       = {10.1142/6821},
  publisher = {{Singapore: World Scientific}}
}

@article{10.1093/ptep/ptaf119,
    author = {Yoshida, Kazuki and Tanaka, Junki},
    title = {Reaction Mechanism of Quasi-Free Knockout Processes in the Exotic RI Beam Era},
    journal = {Progress of Theoretical and Experimental Physics},
    volume = {2026},
    number = {4},
    pages = {04A108},
    year = {2026},
    month = {04},
    issn = {2050-3911},
    doi = {10.1093/ptep/ptaf119}
}

@article{Kramer2001,
 author={Kramer, G. J. and Blok, H. P. and Lapik{\'a}s, L.}, title={A Consistent Analysis of $(e,e'p)$ and $(d,{}^3\mathrm{He})$ Experiments}, journal={Nuclear Physics A}, volume={679}, pages={267--286}, year={2001}, doi={10.1016/S0375-9474(00)00379-1}}

@article{Lapikas1993,
 author={Lapik{\'a}s, L.}, title={Quasi-Elastic Electron Scattering off Nuclei}, journal={Nuclear Physics A}, volume={553}, pages={297c--308c}, year={1993}, doi={10.1016/0375-9474(93)90630-G}}

@incollection{Kelly1996,
 author={Kelly, J. J.}, title={Nucleon Knockout by Intermediate-Energy Electrons}, booktitle={Advances in Nuclear Physics}, volume={23}, pages={75--294}, year={1996}, publisher={Springer}}

@article{Mougey1976,
 author={Mougey, J. and others}, title={Quasi-free (e,e'p) scattering on \(^{12}\){C}, \(^{28}\){Si}, \(^{40}\){Ca} and \(^{58}\){N}}, journal={Nuclear Physics A}, volume={262}, pages={461--492}, year={1976},   doi     = {10.1016/0375-9474(76)90510-8}}

@article{Frick2004,
 author={Frick, T. and Hassaneen, Kh. S. A. and Rohe, D. and M{\"u}ther, H.}, title={Spectral Function at High Missing Energies and Momenta}, journal={Physical Review C}, volume={70}, pages={024309}, year={2004}, doi={10.1103/PhysRevC.70.024309}}

@article{Rios2009,
 author={Rios, A. and Polls, A. and Dickhoff, W. H.}, title={Depletion of the Nuclear Fermi Sea}, journal={Physical Review C}, volume={79}, pages={064308}, year={2009}, doi={10.1103/PhysRevC.79.064308}}

@article{Otsuka2010,
 author={Otsuka, T. and Suzuki, T. and Holt, J. D. and Schwenk, A. and Akaishi, Y.}, title={Three-Body Forces and the Limit of Oxygen Isotopes}, journal={Physical Review Letters}, volume={105}, pages={032501}, year={2010}, doi={10.1103/PhysRevLett.105.032501}}

@article{Hagen2012,
 author={Hagen, G. and Hjorth-Jensen, M. and Jansen, G. R. and Machleidt, R. and Papenbrock, T.}, title={Evolution of Shell Structure in Neutron-Rich Calcium Isotopes}, journal={Physical Review Letters}, volume={109}, pages={032502}, year={2012}, doi={10.1103/PhysRevLett.109.032502}}

@article{Hergert2017,
 author={Hergert, H.}, title={In-Medium Similarity Renormalization Group for Closed and Open-Shell Nuclei}, journal={Physica Scripta}, volume={92}, pages={023002}, year={2017}, doi={10.1088/1402-4896/92/2/023002}}

@article{Soma2013,
 author={Som{\`a}, V. and Barbieri, C. and Duguet, T.}, title={Ab Initio {G}orkov-{G}reen's Function Calculations of Open-Shell Nuclei}, journal={Physical Review C}, volume={87}, pages={011303(R)}, year={2013}, doi={10.1103/PhysRevC.87.011303}}

@article{Ekstrom2015,
 author={Ekstr{\"o}m, A. and others}, title={Accurate Nuclear Radii and Binding Energies from a Chiral Interaction}, journal={Physical Review C}, volume={91}, pages={051301(R)}, year={2015}, doi={10.1103/PhysRevC.91.051301}}

@article{MachleidtEntem2011,
 author={Machleidt, R. and Entem, D. R.}, title={Chiral Effective Field Theory and Nuclear Forces}, journal={Physics Reports}, volume={503}, pages={1--75}, year={2011}, doi={10.1016/j.physrep.2011.02.001}}

@article{Epelbaum2009,
 author={Epelbaum, E. and Hammer, H.-W. and Mei{\ss}ner, U.-G.}, title={Modern Theory of Nuclear Forces}, journal={Reviews of Modern Physics}, volume={81}, pages={1773--1825}, year={2009}, doi={10.1103/RevModPhys.81.1773}}

@article{FurnstahlSchwenk2010,
 author={Furnstahl, R. J. and Schwenk, A.}, title={How Should One Formulate, Extract, and Interpret `Non-Observables' for Nuclei?}, journal={Journal of Physics G: Nuclear and Particle Physics}, volume={37}, pages={064005}, year={2010}, doi={10.1088/0954-3899/37/6/064005}}

@article{Furnstahl2015,
 author={Furnstahl, R. J. and Phillips, D. R. and Wesolowski, S.}, title={A Recipe for {EFT} Uncertainty Quantification in Nuclear Physics}, journal={Journal of Physics G: Nuclear and Particle Physics}, volume={42}, pages={034028}, year={2015}, doi={10.1088/0954-3899/42/3/034028}}

@article{Hebborn2023,
 author={Hebborn, C.  and others}, title={Optical Potentials for the Rare-Isotope Era}, journal={Journal of Physics G: Nuclear and Particle Physics}, volume={50}, pages={060501}, year={2023}, doi={10.1088/1361-6471/acc348}}

@article{Barranco2001,
 author={Barranco, F. and Broglia, R. A. and Col{\`o}, G. and Vigezzi, E. and Bortignon, P. F.}, title={The halo of the exotic nucleus \(^{11}\){L}I: a single {C}ooper pair}, journal={European Physical Journal A}, volume={11}, pages={385--392},   doi = {10.1007/s100500170050}, year={2001}}

@article{CiofiSimula1996,
 author={Ciofi degli Atti, C. and Simula, S.}, title={Realistic Model of the Nucleon Spectral Function in Few- and Many-Nucleon Systems}, journal={Physical Review C}, volume={53}, pages={1689--1710}, year={1996}, doi={10.1103/PhysRevC.53.1689}}

@article{Drischler2021,
 author={Drischler, C. and Holt, J. W. and Wellenhofer, C.}, title={Chiral Effective Field Theory and the High-Density Nuclear Equation of State}, journal={Annual Review of Nuclear and Particle Science}, volume={71}, pages={403--432}, year={2021}, doi={10.1146/annurev-nucl-102419-041903}}

@article{Bogner2010,
 author={Bogner, S. K. and Furnstahl, R. J. and Schwenk, A.}, title={From Low-Momentum Interactions to Nuclear Structure}, journal={Progress in Particle and Nuclear Physics}, volume={65}, pages={94--147}, year={2010}, doi={10.1016/j.ppnp.2010.03.001}}

@article{Carlson2015,
 author={Carlson, J. and others}, title={Quantum Monte Carlo Methods for Nuclear Physics}, journal={Reviews of Modern Physics}, volume={87}, pages={1067--1118}, year={2015}, doi={10.1103/RevModPhys.87.1067}}

@article{Hagen2014,
 author={Hagen, G. and Papenbrock, T. and Hjorth-Jensen, M. and Dean, D. J.}, title={Coupled-Cluster Computations of Atomic Nuclei}, journal={Reports on Progress in Physics}, volume={77}, pages={096302}, year={2014}, doi={10.1088/0034-4885/77/9/096302}}

@article{Hergert2020,
 author={Hergert, H.}, title={A Guided Tour of Ab Initio Nuclear Many-Body Theory}, journal={Frontiers in Physics}, volume={8}, pages={379}, year={2020}, doi={10.3389/fphy.2020.00379}}

@article{Bazin2003,
 author={Bazin, D. and others}, title={New Direct Reaction: Two-Proton Knockout from Neutron-Rich Nuclei}, journal={Physical Review Letters}, volume={91}, pages={012501}, year={2003}, doi={10.1103/PhysRevLett.91.012501}}

@article{Yoneda2006,
 author={Yoneda, K. and others}, title={Two-Proton Knockout from $^{32}${M}g: Intruder Structure of $^{30}${N}e}, journal={Physical Review C}, volume={74}, pages={021303(R)}, year={2006}, doi={10.1103/PhysRevC.74.021303}}

@article{Kobayashi2012,
 author={Kobayashi, N. and Nakamura, T. and Kondo, Y. and others}, title={One-Neutron Removal from $^{31}$Ne: Evidence for a $p$-Wave Halo}, journal={Physical Review C}, volume={86}, pages={054604}, year={2012}, doi={10.1103/PhysRevC.86.054604}}

@article{Aumann2000,
 author={Aumann, T. and others}, title={One-Neutron Knockout from Individual Single-Particle States of $^{11}${Be}}, journal={Physical Review Letters}, volume={84}, pages={35--38}, year={2000}, doi={10.1103/PhysRevLett.84.35}}

@article{Sauvan2004,
 author={Sauvan, E. and others}, title={One-Neutron Removal Reactions on Neutron-Rich $p$-$sd$ Shell Nuclei}, journal={Physical Review C}, volume={69}, pages={044603}, year={2004}, doi={10.1103/PhysRevC.69.044603}}

@article{Gade2004,
  title = {Knockout from $^{46}\mathrm{Ar}:\ensuremath{\ell}=3$ neutron removal and deviations from eikonal theory},
  author = {Gade, A. and others},
  journal = {Phys. Rev. C},
  volume = {71},
  issue = {5},
  pages = {051301(R)},
  numpages = {5},
  year = {2005},
  month = {May},
  publisher = {American Physical Society},
  doi = {10.1103/PhysRevC.71.051301}
}

@article{Rohe2004,
 author={Rohe, D. and others}, title={Correlated Strength in the Nuclear Spectral Function}, journal={Physical Review Letters}, volume={93}, pages={182501}, year={2004}, doi={10.1103/PhysRevLett.93.182501}}

@article{Sargsian2014,
 author={Sargsian, M. M.}, title={New Properties of the High-Momentum Distribution of Nucleons in Asymmetric Nuclei}, journal={Physical Review C}, volume={89}, pages={034305}, year={2014}, doi={10.1103/PhysRevC.89.034305}}

@article{Ryckebusch2019,
 author={Ryckebusch, J. and Cosyn, W. and Stevens, S. and Casert, C. and Nys, J.}, title={The isospin and neutron-to-proton excess dependence of short-range correlations}, journal={Physics Letters B}, volume={792}, pages={21--28}, year={2019},  doi = {10.1016/j.physletb.2019.03.016}}

@article{Weinstein2011,
 author={Weinstein, L. B. and others}, title={Short Range Correlations and the {EMC} Effect}, journal={Physical Review Letters}, volume={106}, pages={052301}, year={2011}, doi={10.1103/PhysRevLett.106.052301}}

@article{Duer2018,
 author={Duer, M. and Hen, O. and Piasetzky, E. and others}, title={Probing High-Momentum Protons and Neutrons in Neutron-Rich Nuclei}, journal={Nature}, volume={560}, pages={617--621}, year={2018}, doi={10.1038/s41586-018-0400-z}}

@article{Alvioli2005,
 author={Alvioli, M. and Ciofi degli Atti, C. and Morita, H.}, title={Ground-State Energies, Densities and Momentum Distributions in Closed-Shell Nuclei Calculated within a Cluster Expansion Approach and Realistic Interactions}, journal={Physical Review C}, volume={72}, pages={054310}, year={2005}, doi={10.1103/PhysRevC.72.054310}}

\end{document}